# Current Methods for Hyperpolarized [1-$^{13}$C]pyruvate MRI Human Studies

Authors:
Peder EZ Larson [1, 2,*], Jenna ML Bernard[1], James A Bankson [3], Nikolaj Bøgh [4], Robert A Bok[1], Albert P. Chen [5], Charles H Cunningham [6,7], Jeremy Gordon[1], Jan-Bernd Hövener [8], Christoffer Laustsen [4], Dirk Mayer [9,10], Mary A McLean[11 12], Franz Schilling[13], James Slater[1], Jean-Luc Vanderheyden[5, 14], Cornelius von Morze [15], Daniel B Vigneron[1, 2], Duan Xu[1, 2], and the HP 13C MRI Consensus Group[#]

[1]Department of Radiology and Biomedical Imaging, University of California, San Francisco, CA 94143, USA.
[2]UC Berkeley-UCSF Graduate Program in Bioengineering, University of California, San Francisco and University of California, Berkeley, CA 94143, USA.
[3] Department of Imaging Physics, MD Anderson Medical Center, Houston, TX, USA.
[4] The MR Research Center, Department of Clinical Medicine, Aarhus University, Aarhus, Denmark.
[5] GE Healthcare, Menlo Park, California, USA.
[6] Physical Sciences, Sunnybrook Research Institute, Toronto, Ontario, Canada.
[7] Department of Medical Biophysics, University of Toronto, Toronto, Ontario, Canada.
[8] Section Biomedical Imaging, Molecular Imaging North Competence Center (MOIN CC), Department of Radiology and Neuroradiology, University Medical Center Schleswig-Holstein (UKSH), Kiel University, Am Botanischen Garten 14, 24118, Kiel, Germany.
[9] Department of Diagnostic Radiology and Nuclear Medicine, University of Maryland School of Medicine, Baltimore, MD, USA.
[10] Greenebaum Cancer Center, University of Maryland School of Medicine, Baltimore, MD, USA
[11] Department of Radiology, University of Cambridge, Cambridge, United Kingdom.
[12] Cancer Research UK Cambridge Institute, University of Cambridge, Li Ka Shing Center, Cambridge, United Kingdom.
[13] Department of Nuclear Medicine, School of Medicine, Klinikum Rechts der Isar, Technical University of Munich, 81675 Munich, Germany.
[14]JLVMI Consulting LLC, Dousman, WI, USA
[15] Department of Radiology, Washington University, St. Louis, MO, USA.
[#]See Acknowledgements for a list of all HP $^{13}$C MRI Consensus Group Members
[*]Corresponding Author: Peder Larson peder.larson@ucsf.edu

This work was supported by the ISMRM Hyperpolarized Media MR Study Group, the ISMRM Hyperpolarization Methods & Equipment Study Group, and the Hyperpolarized MRI Technology Resource Center (NIH/NIBIB grant P41EB013598).

# Abstract

MRI with hyperpolarized (HP) $^{13}C$ agents, also known as HP $^{13}C$ MRI, can measure processes such as localized metabolism that is altered in numerous cancers, liver, heart, kidney diseases, and more.  It has been translated into human studies during the past 10 years, with recent rapid growth in studies largely based on increasing availability of hyperpolarized agent preparation methods suitable for use in humans. This paper aims to capture the current successful practices for HP MRI human studies with [1-$^{13}C$]pyruvate - by far the most commonly used agent, which sits at a key metabolic junction in glycolysis.  The paper is divided into four major topic areas: (1) HP $^{13}C$-pyruvate preparation, (2) MRI system setup and calibrations, (3) data acquisition and image reconstruction, and (4) data analysis and quantification. In each area, we identified the key components for a successful study, summarized both published studies and current practices, and discuss evidence gaps, strengths, and limitations. This paper is the output of the "HP $^{13}C$ MRI Consensus Group" as well as the ISMRM Hyperpolarized Media MR and Hyperpolarized Methods & Equipment study groups. It further aims to provide a comprehensive reference for future consensus building as the field continues to advance human studies with this metabolic imaging modality.

# Introduction

MRI with hyperpolarized $^{13}C$ agents, also known as hyperpolarized (HP) $^{13}C$ MRI, has shown great potential as a novel imaging modality, particularly for its ability to probe metabolic processes in real time. The first human studies with HP [1-$^{13}C$]pyruvate were performed in 2011 in prostate cancer patients (1).   Since then, there have been over 60 papers published with imaging results of human subjects from 13 different sites, with applications including prostate cancer, brain tumors, breast cancer, kidney cancer, pancreatic cancer, metastatic disease, liver disease, ischemic heart disease, diabetes and cardiomyopathies.  The vast majority of these studies used [1-$^{13}C$]pyruvate (1–63), where [2-$^{13}C$]pyruvate (64) and $^{13}C$-urea (56) have been demonstrated too.

As clinical HP $^{13}C$ MRI advances, there is a growing need to build consensus for best practices, which are critical for comparing data across sites, performing multi-site trials,deploying methods to new sites, partnering with vendors, and potentially for obtaining broader regulatory approvals. In March 2022, we initiated an effort to build consensus within the HP $^{13}C$ MRI community with this opportunity in mind, and it was greeted with strong enthusiasm.

The "HP $^{13}C$ MRI Consensus Group", containing over 55 members from 27 sites, identified the area of greatest need and opportunity for consensus building to be HP [1-$^{13}C$]pyruvate human studies, for the following reasons:

- Pyruvate is the most mature and widely used HP agent and has the most significant translational evidence emphasizing the potential clinical impact.
- Clinical trials, particularly multi-site trials, have the strongest need for consensus methods to ensure that data can be combined across sites.

This work is a Position Paper for which the goal is to describe current successful practices and study methods for HP [1-$^{13}$C]pyruvate human studies along with justification to support those practices. This is divided into four major topic areas: (1) HP $^{13}$C-pyruvate preparation, (2) MRI system setup and calibrations, (3) data acquisition and image reconstruction, and (4) data analysis and quantification (Fig. 1). The current successful practices and study methods include a literature review of published peer-reviewed journal papers showing human HP [1-$^{13}$C]pyruvate study data, up to September 2022 (1–63), as well as new unpublished information from surveys of HP $^{13}$C study sites. Based on this information, we also highlight the evidence gaps, strengths, and limitations of current practices which are summarized at the end of each section.

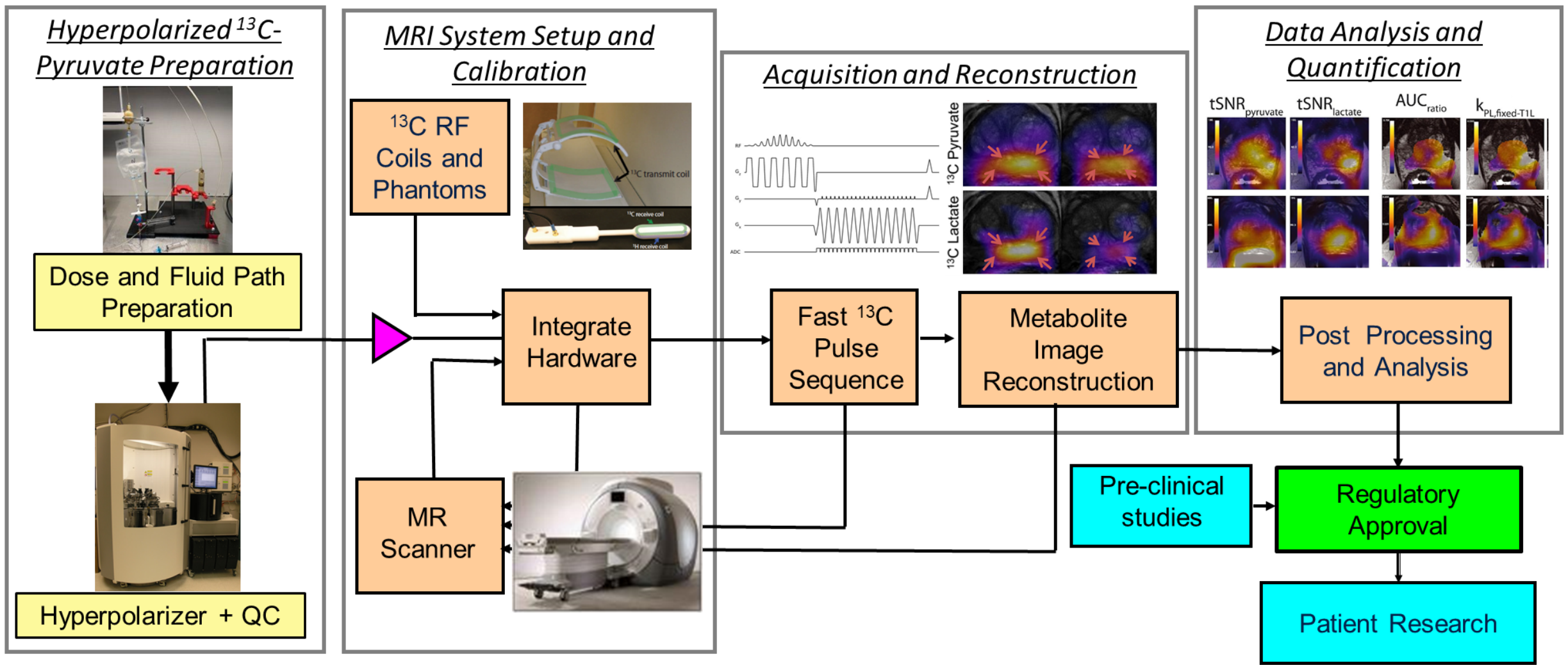

**Figure 1**: Illustration of the HP $^{13}$C MRI human study process, including the 4 major areas covered in this paper: Hyperpolarized $^{13}$C-pyruvate preparation, MRI system setup and calibration, Acquisition and Reconstruction, and Data Analysis and Quantification.

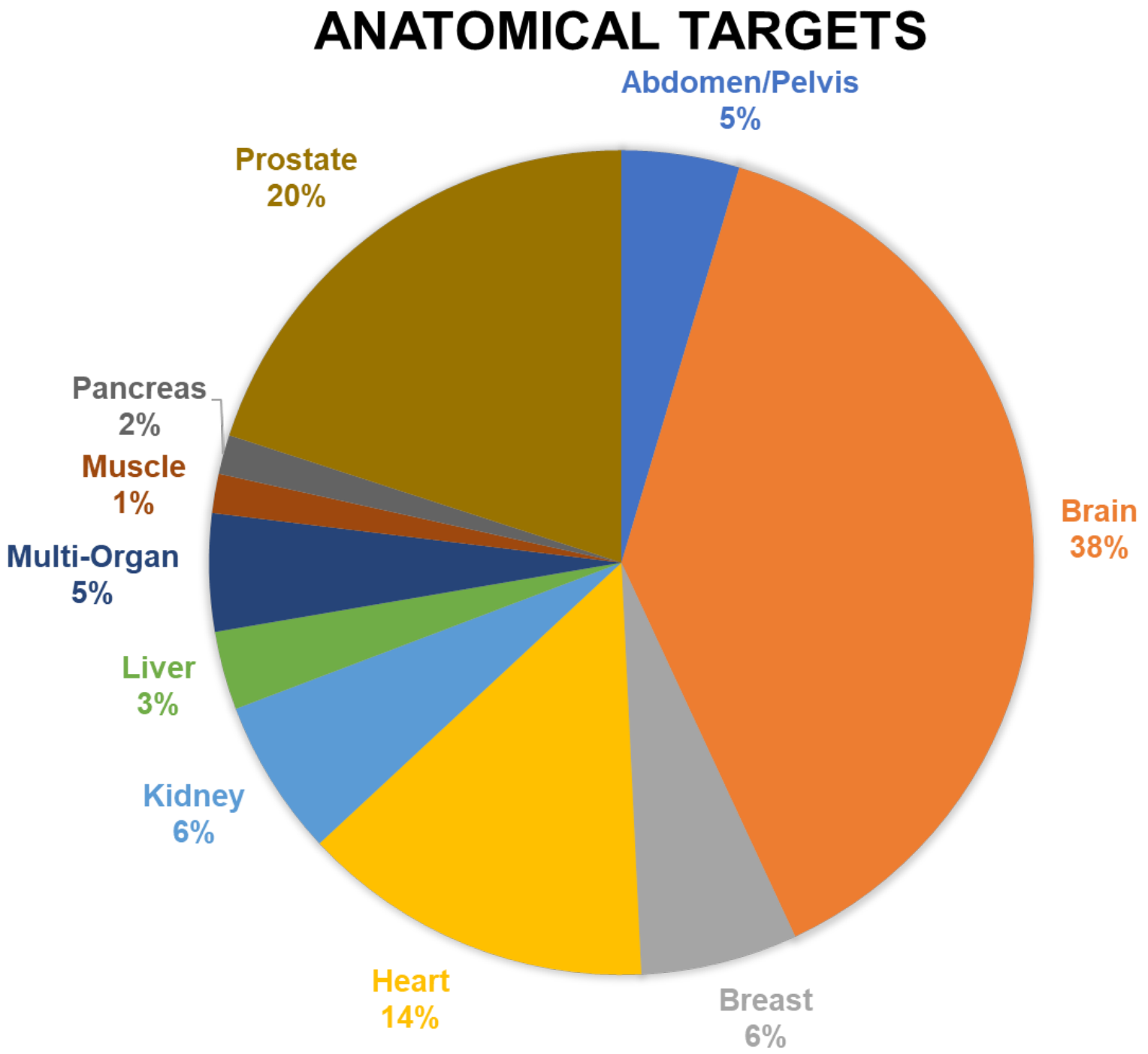

**Figure 2**: Anatomical targets of HP [1-$^{13}$C]pyruvate MRI human studies published up to September 2022.

# Hyperpolarized $^{13}$C-Pyruvate Preparation

This section covers the processes for creating the HP agent, $^{13}$C pyruvate, and will include many aspects and considerations that are needed to safely and effectively prepare doses for metabolic imaging studies in human subjects. These include material, personnel, equipment and facility, fluid path preparation, quality control, and release.

It is helpful to understand that the specifications of a dose of $^{13}$C pyruvate suitable for in vivo MR HP metabolic imaging were shaped in part by early preclinical studies performed by GE HealthCare summarized in Ref. (65). In short, the safety of the two novel drug components, $^{13}$C pyruvate and the electron paramagnetic agent (EPA) AH111501, were demonstrated in those studies. The more precise formulation of the dose suitable for human use was then determined from clinical studies (66) that included two Phase 1 clinical trials in young and elderly healthy volunteers without hyperpolarization of the $^{13}$C nuclei and another Phase 1/2a dose escalation and imaging feasibility study with HP $^{13}$C pyruvate in 31 prostate cancer patients at the University of California, San Francisco (UCSF) (1).

With the exception of the first HP $^{13}$C imaging clinical trial, which utilized a prototype device in a cleanroom (1), all HP $^{13}$C studies performed in humans to date have utilized the SPINlab polarizer (manufactured by GE HealthCare). Consequently all doses of the HP $^{13}$C pyruvate delivered by SPINlab have been produced using the “SPINlab Pharmacy Kit” that serves as the container-closure system for the various drug components ($^{13}$C pyruvic acid and EPA mixture, dissolution medium, and neutralization and dilution medium) during sample polarization, dissolution and quality control (QC) processes. Thus many aspects of the HP sample preparation considerations discussed below are related to the SPINlab instrument and the consumables designed to be used with it (67).

## General considerations

While more than 860 patients or healthy subjects having been injected with HP $^{13}C$ pyruvate as of January 2022 without reports of any serious adverse events (68), HP $^{13}C$ pyruvate injection remains an investigational MR contrast agent and can only be administered by those with Investigational New Drug (IND) exemption from the Food and Drug Administration (FDA) in the USA, a Clinical Trial Application (CTA) in Canada, approval from National Research Ethics Committee Services in the UK, or approval from the relevant local regulatory body. Thus, methods and processes involved to produce a dose should have patient safety as the first priority. Since utilizing dissolution dynamic nuclear polarization (dissolution-DNP) for human use is still a relatively new development, there are no existing published regulatory guidelines specifically for this method.

There are two major production styles that determine how various sites approach the agent preparation. In the US, the most common approach is to rely on a sterilizing filter ("Terminal Sterilization") to ensure sterility of the final product, akin to PET tracer production, where a starting molecule with a radioisotope is processed using various other ingredients to make the final, desired and injectable contrast agent within a necessarily short amount of time (69). For these sites, sterilization of the components and accessories upstream of this filter are not required, although many of them were manufactured and tested following Good Manufacturing Practice (GMP) or Good Laboratory Practice (GLP) requirements. The filling process is usually performed under an ISO 5 laminar flow hood, but a clean room or an isolator is not required. This approach is typically accompanied by testing the integrity of the sterilizing filter prior to release of the dose for injection. Typically, post release endotoxin and sterility tests are performed using an aliquot reserved from each released dose.

In the UK and EU, the most common approach is to more-closely follow sterile pharmaceutical compounding guidelines (70), where all components and ingredients are required to be sterile or manufactured under GMP guidelines and are assembled and filled within a clean room environment or an isolator system ("Sterile Preparation"). Typically a batch of Pharmacy Kits for HP $^{13}C$ pyruvate injection are prepared together. The sterility of the final dose is also ensured by batch validation testing, in addition to the sterility of the ingredients and the sterile compounding process. The endotoxin and sterility testing are performed for the process validation but are not performed for each injected dose. Some institutions fill and assemble the Pharmacy Kit required for a specific study on the same day or the day prior to polarization, dissolution, and patient administration, but others have also demonstrated the feasibility of preparing a batch of kits, keeping them in a -20ºC freezer and using them over a period of a few months.

Beyond the obvious requirements that the process and the facility has to ultimately produce a dose that is safe to inject into a human, regulatory authorities will also focus on the question “Are you in control of your processes?”. To be in control of your process requires an in-depth and broad understanding of all processes involved in pre, post, and during the production process.

## Personnel

It is typical and may be required to have licensed personnel involved in the production process depending on local regulations.Typically a pharmacist, radiopharmacist or other similarly qualified person (QP), in charge of the facility where the Pharmacy Kit filling and preparation is taking place, is responsible for the overall process and the release of the injectable dose. Qualified cleanroom technicians are often involved in the Pharmacy Kit filling under the supervision of the pharmacist or QP. As is required for pharmaceutical compounding or PET tracer production, training requirements and training records for all personnel need to be maintained and available for audit by the FDA or equivalent.

## Equipment and facility

The facility and all equipment need to have standard operating procedures (SOPs) that describe how equipment is used, maintained, and calibrated to comply with relevant legislation. Currently, almost all the filling of the Pharmacy Kit takes place within a compounding laminar flow hood or isolator (typically ISO 5). At some sites, the filling is conducted within a cleanroom, while at others, it is conducted in a dedicated non-cleanroom space, reflecting differences in cleanroom approach and specifications between regulators worldwide (71). Some equipment or facilities, such as the compounding hood or cleanroom, may require external certified laboratories for testing.

## Material handling

Material handling guidelines (69,70) require SOPs detailing a system to track all of the materials involved in the HP production process for a particular patient dose, similar to current good manufacturing practice (cGMP) requirements for material handling for drug compounding. This includes acceptance standards, storage conditions, amount used in the patient dose for each ingredient and materials used in the assembly of the fluid path and Pharmacy Kit. Currently some users choose to open and inspect and sometimes modify the Pharmacy Kits upon arrival, but some users keep them in the sealed packaging until they are required for dose preparation.

## Pharmacy Kit filling and assembling

As required by an IND or its equivalent, the preparation of the doses of HP $^{13}C$ agent are detailed in the Chemistry, Manufacturing, and Control (CMC) section of an applicable regulatory submission; an example of this has been made available (72). It describes the processes of filling the Pharmacy Kit with the different components that make up the final drug product, and of assembling the final kit for either storage or immediate use in the polarizer. Special attention should be given to the laser welding process in order to satisfy installation qualification (IQ) and operational qualification (OQ). Typically, the final developed process is validated by process qualification (PQ) runs, during which 3 or more Pharmacy Kits are filled and used and the final HP $^{13}C$ products are tested for endotoxin and sterility and to confirm that they meet the dose specifications for injections (usually including pyruvate concentration, residual EPA concentration, pH, liquid state polarization level and dose temperature). The data from 3 consecutive PQ runs are submitted as part of the IND submission (or its equivalent), and are often also reviewed by the Institutional Review Board (IRB) where the studies are conducted.

## Quality control and dose release

The quality control (QC) and dose release can be separated into two aspects: one is the QC and release of the filled Pharmacy Kit, and second is the QC and release of the HP $^{13}C$ agent for injection, after polarization and dissolution. For institutions filling a batch of kits and storing them to use over a period of time, typically the batch can be released based on initial validation, environmental monitoring data from the day of kit production, and if filters are used during preparation of any of the components, filter integrity testing. But in some cases one or more kits are used for validation before the batch of kits are released for future use. For institutions that fill only the kits required for specific studies shortly before the experiment, the filled kits often do not go through separate release tests before they are used.

The quality control of the HP $^{13}C$ pyruvate solution post dissolution is primarily performed to ensure that the agent meets the dose specifications (Table 1) before it is administered to the subject. These specifications target both safety (pH, residual EPA, temperature) and efficacy (pyruvate concentration, polarization, volume). Typically, the pyruvate concentration, residual EPA concentration, pH, dose temperature, dose volume, and liquid state polarization are measured by the QC accessory associated with the SPINlab polarizer. Some users perform a secondary measurement for one of the parameters, such as pH, using a different instrument or pH paper. For sites that do not go through a separate release testing process for batch filled kits, the integrity of the sterilization assurance filter, a part of the Pharmacy Kit, is typically tested as a part of the dose release. It is also common for these users to preserve an aliquot of the final HP $^{13}C$ pyruvate solution for post-release endotoxin and sterility testing. This testing cannot be completed fast enough to test an individual dose prior to injection, but this is why other processes such as PQ runs and validation testing are done to minimize the chance a subject could be injected with a contaminated dose. The final dose release and injection should be done under the supervision of a licensed professional, based on local regulations.

## Some key challenges

Many of the challenges associated with HP $^{13}C$ pyruvate preparation can be attributed to the conditions required for the dissolution-DNP method of high magnetic field (~3-7 T) and very low temperature (~1 K) during polarization, with pressurized and superheated water necessary for the rapid dissolution event. These extreme conditions are quite challenging for the design of the container-closure and fluid path system. In particular, the cryogenic temperature in the polarizer requires special attention to any moisture or ambient (moist) air introduced into that portion of the fluid path, which can form an ice block at ~1 K. This ice can lead to flow restriction during the dissolution event and reduce the strength of the laser welded bond between the cryovial and its cap. This can ultimately produce failures in the dissolution step, including variations in final pyruvate concentration and pH that may fail to meet QC release criteria as well as fluid path ruptures that provide no available dose and result in polarizer down-time.

The polarization of the HP $^{13}C$ pyruvate sample decays quickly over the span of a few minutes after dissolution, and thus the process of dissolution, QC for release, and injection should be completed as fast as possible to preserve the high polarization level achieved. Any delays in the preparation process, such as transportation time or equipment malfunction, can significantly reduce the final polarization and result in lower quality imaging data.

## Current Practices

A summary of data collected from all sites performing clinical trials with HP $^{13}C$-pyruvate is shown in Fig. 3 and Table 1, including the specification of the final dose and how the quality control and release of the final dose are performed. There is a split in the Production Style, described in the General Considerations section above, with 8/13 sites using Sterile Preparation versus 5/13 using Terminal Sterilization. While many of the dose specifications show notable differences in acceptable ranges, all of these variations listed in tables have been successfully and safely been used to perform HP $^{13}C$ pyruvate studies in humans. Their differences depend on the institutions' preferences, resources and their particular regulatory situation. There is high similarity in pyruvate ranges, temperature ranges, EPA limits, and volume limits. There is modest variability in pH ranges and large variability in the endotoxin test limit. There is a 3-fold difference in acceptable polarization levels, which are measured to ensure a futile dose is not injected since the polarization is directly proportional to SNR. This reflects the decision by several sites to believe that useful data can be still be obtained with suboptimal polarizations.

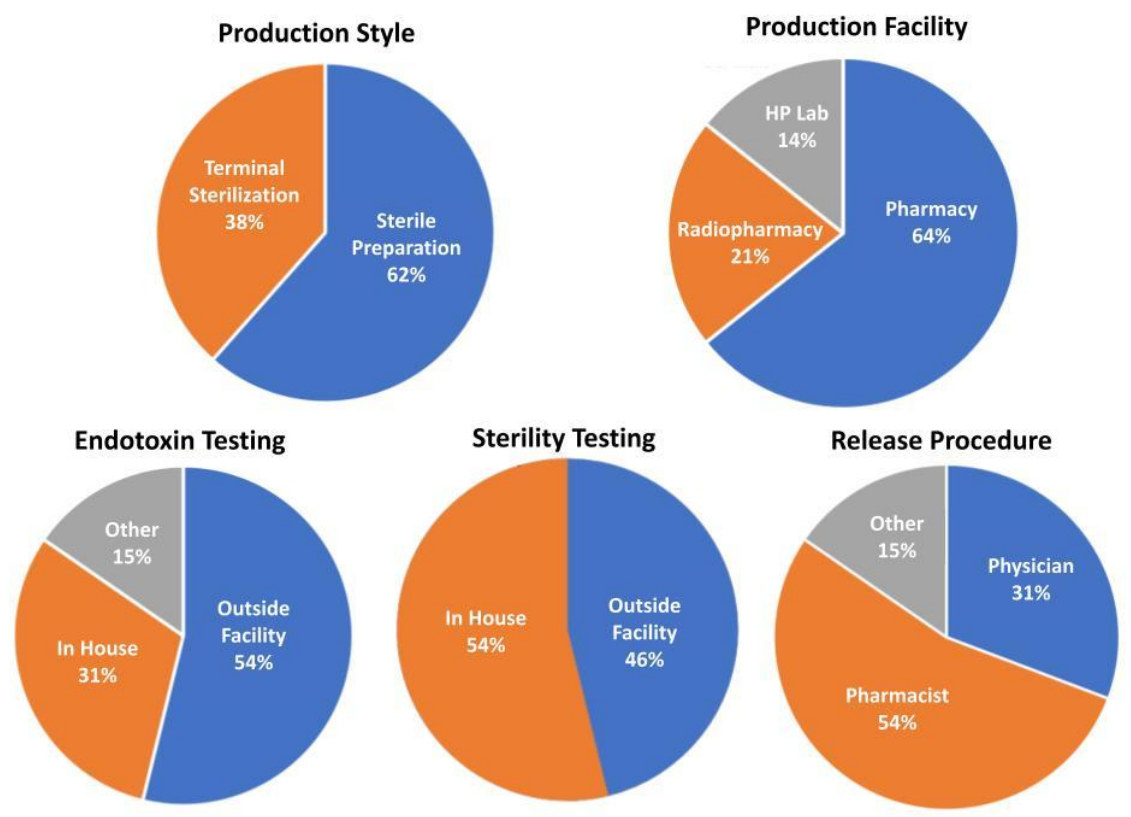

**Figure 3**: Hyperpolarized agent preparation methods reported by sites currently performing HP $^{13}$C MRI human studies

| Site # | Production Style | Facility | Release Procedure | pH | Pyruvate [mM] | Residual EPA [mM] | Polarization, Liquid state [%] | Temperature [°C] | Volume [mL] | Appearance test performed | Filter Integrity Test | Endotoxin Test [EU/mL] | Endotoxin Testing | Sterility Test | Sterility Testing |
|---|---|---|---|---|---|---|---|---|---|---|---|---|---|---|---|
| 1 | Sterile Preparation | Pharmacy | Other | **4 - 9** | **200 - 300** | **<2.6** | **> 7.5** | **20 – 42** | **≥35** | Yes | **Yes** | <3.7 | Outside | No growth | Outside |
| 2 | Sterile Preparation | Radiopharmacy | Physician | **5 - 9** | **200 - 280** | **<5.0** | **>5.0** | **25 - 37** | **≥40** | **Yes** | No | <0.02 | Outside | No growth | In House |
| 3 | Sterile Preparation | Pharmacy | Pharmacist | **6.7- 8.3** | N/A | **<3.0** | **≥15** | **25 – 40** | **>38** | **Yes** | No | <0.125 | Outside | No growth | Outside |
| 4 | Terminal Sterilization | Hyperpolarized | Other | **6.5- 8.5** | **150 - 280** | **<3.0** | NA | 25 – 40 | 6 ml + patient dose | NA | **Yes** | <5.0 | Other | No growth | In House |
| 5 | Sterile Preparation | Pharmacy | Pharmacist | **6.7– 8.2** | **220 - 280** | **<3.0** | **≥5** | **25 – 39** | **>40** | **Yes** | No | <11 | Outside | No growth | Outside |
| 6 | Sterile Preparation | Pharmacy | Pharmacist | **6.5-8.2** | **210 - 280** | **<3.0** | ≥15 | **25 – 39** | **>40** | **Yes** | **Yes** | <0.5 | Outside | No growth | Outside |
| 7 | Terminal Sterilization | Pharmacy | Pharmacist | **6.5-8.5** | **200 - 280** | **<3.0** | **≥15** | **25 – 40** | **≥38** | Yes | **Yes** | <4.37 | In House | No growth | In House |
| 8 | Sterile Preparation | Pharmacy | Physician | **6.5-8.9** | **220 - 280** | **<3.0** | **≥15** | **25 – 37** | **≥40** | **Yes** | No | <0.5 | Outside | No growth | Outside |
| 9 | Terminal Sterilization | Hyperpolarized | Pharmacist | **5 - 9** | **185 - 280** | **<5.0** | NA | 25 – 40 | **>38** | **Yes** | **Yes** | <2.5 | In House | No growth | In House |
| 10 | Sterile Preparation | Pharmacy | Physician | **6.5-8.5** | **220 - 280** | **<3.0** | ≥10 | **25 – 37** | **>38** | **Yes** | No | <0.1 | Outside | No growth | Outside |
| 11 | Terminal Sterilization | Pharmacy | Physician | **6.5-8.5** | **210 - 280** | **<3.0** | **≥15** | **25 – 37.4** | **>40** | **Yes** | **Yes** | NA | Other | No growth | In House |
| 12 | Terminal Sterilization | Radiopharmacy | Pharmacist | **6.7-8.6** | **220 - 280** | **<3.0** | **≥15** | **25 – 37** | **>40** | **Yes** | **Yes** | <5.0 | In House | No growth | In House |
| 13 | Sterile Preparation | Pharmacy /Radiopharmacy | Pharmacist | **5 - 9** | **220 - 280** | **<3.0** | **≥15** | **25 – 37** | **>38** | NA | **Yes** | NA | In House | No growth | In House |

**Table 1**: HP $^{13}$C-pyruvate preparation parameters, methods, and dose specifications used for quality control testing and release as well as validation. These were obtained from a survey of all sites performing clinical trials with HP [1-$^{13}$C]pyruvate. The parameters used for product release are noted in **bold text**, otherwise these parameters are measured for batch validation or other QC measurements. The endotoxin and sterility testing are performed during process validation of the batch and/or post-injection, and largely depends on the agent production approach.

## Summary

The overall safety record of HP $^{13}$C-pyruvate has been very strong, and the SPINlab hyperpolarizer has proven to provide high polarizations at human sized doses while meeting numerous QC and release criteria. A weakness remains the failure modes of the SPINlab Phamacy Kits (e.g. ice blocks, path ruptures), which are placed under extreme requirements particularly during dissolution. The preparation process still requires a high degree of expertise. Therefore, there is a significant need to improve the reliability, robustness, and ease of operation for generating HP $^{13}$C-pyruvate doses for human studies. Furthermore, there is a divide between manufacturing and sterile compounding style preparation as well as other site-specific practices, resulting in variations in SOPs and justification required to relevant regulatory bodies. There have also been no comparisons between these approaches. It is also unclear what release criteria and QC parameters are truly required to ensure patient safety. However, all of the reported methods are acceptable and approved by the appropriate regulatory authorities, and have led to the rapid expansion of successful human studies in recent years.

# MRI System Setup and Calibrations

This section covers the MRI system setup, including the imaging system, RF coils, phantoms, and prescan calibration methods.

## Imaging system

The main prerequisite for a given MRI scanner to be capable of supporting studies with HP $^{13}C$ is its “broadband” capability to transmit and receive radiofrequency (RF) signal at the frequency of $^{13}C$, which is around 4 times lower than $^{1}H$. This does not come as a default on clinical MR devices. The transmit power of the broadband amplifier should also be sufficient to support the intended flip angle and RF pulse shape with the employed transmission RF coil(s) for $^{13}C$. Most studies to date use relatively low flip angles (< 90 degrees) for HP $^{13}C$ in order to preserve polarization for time-resolved imaging. The capability to receive $^{13}C$ signal on multiple channels is also desirable to increase SNR, as discussed further in the “RF coils” section.

The choice of magnetic field strength is primarily dependent on the metabolites’ frequency separation due to chemical shift dispersion and $^{1}H$ imaging. High field strengths do not enhance hyperpolarized $^{13}C$ signal as they do for $^{1}H$ because the signal strength in a HP experiment relies on manipulating the population of quantum energy states outside of the MRI scanner. However, the injected HP $^{13}C$-pyruvate and its metabolic products have greater frequency separation at higher fields, and it may thus be easier to separate and quantify these resonances at higher fields. This comes at the cost of a reduction in the achievable $T_2$* and often reduced $T_1$. As the initial polarization is independent of the imaging field strength it has been proposed that the increased $T_2$* at 1.5T can potentially be exploited to increase SNR by adapting the acquisition bandwidth or reduce off-resonance imaging effects in cases when the decay of the transverse magnetization is dominated by $T_2$* (73). In practice, 3T has been used in all published human $^{13}C$-pyruvate studies surveyed (Supporting Table S1), and comprises the majority of scanners currently in use for human studies (Table 3). A field strength of 3T is well-suited for $^{1}H$ MRI anatomical reference and correlative imaging.

Stronger and more rapidly slewing magnetic field gradients support more rapid spatial encoding, particularly for metabolite-specific single-shot imaging using echo-planar imaging (EPI) or spiral imaging (See “Acquisition and Reconstruction”). Although the spatial resolution acquired for HP $^{13}C$ imaging is typically much coarser than for $^{1}H$ MRI, the factor of ~4 in gyromagnetic ratio leads to the same reduction factor in performance of the gradient system, so $^{13}C$ experiments are potentially more limited by gradient hardware performance. To date, all human studies have used the commercially-available integrated gradient systems provided in clinical MRI scanners.

Optimization of scanner design has understandably focused on minimization of artifacts in $^{1}H$ MRI, where devices such as room lights, the gradient amplifiers, and the motors driving the patient bed are checked to ensure that they do not produce RF interference at the $^{1}H$ frequency, but artifacts may arise at other frequencies. Eddy current compensation is also not always appropriately adjusted for nuclei at other frequencies (74). In order to optimize for $^{13}C$, many sites have performed checks on phantoms for RF interference, gradient artifacts, and eddy currents (74), including the use of post-hoc gradient impulse response function characterisation and correction, and some vendors have fixed these issues as well.

## RF coils

For HP $^{13}C$ imaging studies in humans, RF coils for both $^{1}H$ and $^{13}C$ nuclei are needed, with $^{1}H$ MRI providing an anatomical reference for registration and optional additional multiparametric MRI readouts. At the Larmor frequency of $^{13}C$ nuclei, the relative contributions from coil noise compared to sample noise increase compared to $^{1}H$ (73,75), although sample noise still is likely the dominant contributor for human-sized coils at 32.1MHz - the resonance frequency of $^{13}C$ nuclei at 3T.

The key requirement for human $^{13}C$-pyruvate RF coils are that the coil geometry and sensitive volume must cover the volume of interest in the subject. Table 2 and Figure 4 shows coil configurations that have been used and optimized for applications in different anatomic regions.

Volume resonators are most commonly used for transmit, as they surround the subject to provide B1 transmit across the FOV ($B_1^+$). While $^{1}H$ relies on a large birdcage ("body") coil built into the scanner, $^{13}C$ transmit coils must be placed inside the bore. This takes up valuable space within the magnet, and also has led to the use of designs with relatively inhomogeneous $B_1^+$. Many human studies have used Helmholz pair resonators for transmit, including the "clamshell coil", which has a notably inhomogeneous $B_1^+$ profile but has been used because of relatively easy integration into the scanner bore. $B_1^+$ variation results in variations in the flip angles that control the use of the hyperpolarized magnetization and creates errors in common HP metrics (9,76). The exception are head coils, where birdcage designs with highly homogeneous $B_1^+$ can be placed around the head while easily fitting inside the bore.

As with $^{1}H$ MRI, higher SNR can typically be achieved by smaller receive coil elements, such as surface coils or phased arrays, and the majority of $^{13}C$ receive coils used have layouts similar to $^{1}H$ phased arrays.

RF coil quality control is important to ensure proper functioning of the coils to provide consistent imaging quality, especially with limited natural abundance $^{13}C$ signal *in vivo*. It typically involves 1) a physical integrity check of the coil cables and connectors and 2) phantom SNR tests to check the coil's performance and to monitor it over time (see Phantoms below). An useful reference for RF coil quality control is outlined in the MRI accreditation program of the American College of Radiology (77) and can be adapted for $^{13}C$ coils.

Notably, configurations for brain and prostate studies used dual-tuned $^{1}H/^{13}C$ coil designs, which greatly simplify workflow and registration of $^{1}H$ and $^{13}C$ images, as no switching of coils is needed.

| Anatomical Target | $^{13}$C TX-transmit coil type | $^{13}$C RX-receive coil type | $^{1}$H coil setup | Representative Reference |
|---|---|---|---|---|
| Abdomen/Pelvis | clamshell coil | 16-channel bilateral array | Body coil | (4) |
| Abdomen/Pelvis | clamshell coil | 2× 4-channel paddle arrays | Body coil or 32-channel abdomen array, repositioned | (37) |
| Abdomen/Pelvis | clamshell coil | 8-channel array | 32-channel abdomen array, repositioned | (46), Fig 4A, B |
| Abdomen/Pelvis | semi-flexible quadrature coil | 2× 4-channel arrays | Body coil | (62) |
| Brain | clamshell coil | 2× 4-channel paddle arrays | 8-channel head array, repositioned | (5) |
| Brain | birdcage coil | 32-channel array | Body coil | (14), Fig 4C |
| Brain | quadrature coil | 8-channel array | quadrature TX/RX head coil | (50) |
| Brain | dual-tuned $^{1}$H/$^{13}$C quadrature coil | dual-tuned $^{1}$H/$^{13}$C quadrature coil | dual-tuned $^{1}$H/$^{13}$C quadrature coil | (55) |
| Brain | clamshell coil | 8-channel receive array | Body coil | (57) |
| Breast | 2-channel TX/RX | 2-channel TX/RX and 6-channel RX-only | Body coil or 8-channel breast array, repositioned | (43) |
| Heart | clamshell coil | 2× 4-channel paddle arrays | not reported | (2) |
| Heart | Helmholtz 2-loop coil-pair | Helmholtz 2-loop coil-pair | Body coil | (36) |
| Heart | semi-flexible quadrature coil | 2× 4-channel arrays | Body coil | (36) |
| Heart | clamshell coil | 16-channel receive array | Body coil or 8-channel cardiac array, repositioned | (52) |
| Prostate | clamshell coil | dual $^{1}$H/$^{13}$C endorectal coil | dual $^{1}$H/$^{13}$C endorectal coil | (1) |

**Table 2**: RF coil configurations reported for human HP [1-$^{13}$C]pyruvate studies. TX = transmit coil, RX = receive coil. The commonly used “clamshell” TX coil is a Helmholz pair design. For $^{1}$H RF configurations, all used the Body coil for TX unless otherwise noted, and “repositioned” indicates the $^{13}$C coil was removed for $^{1}$H imaging. One representative reference is listed for each configuration. The RF coil configurations reported in the reviewed papers are shown in Supporting Table S1.

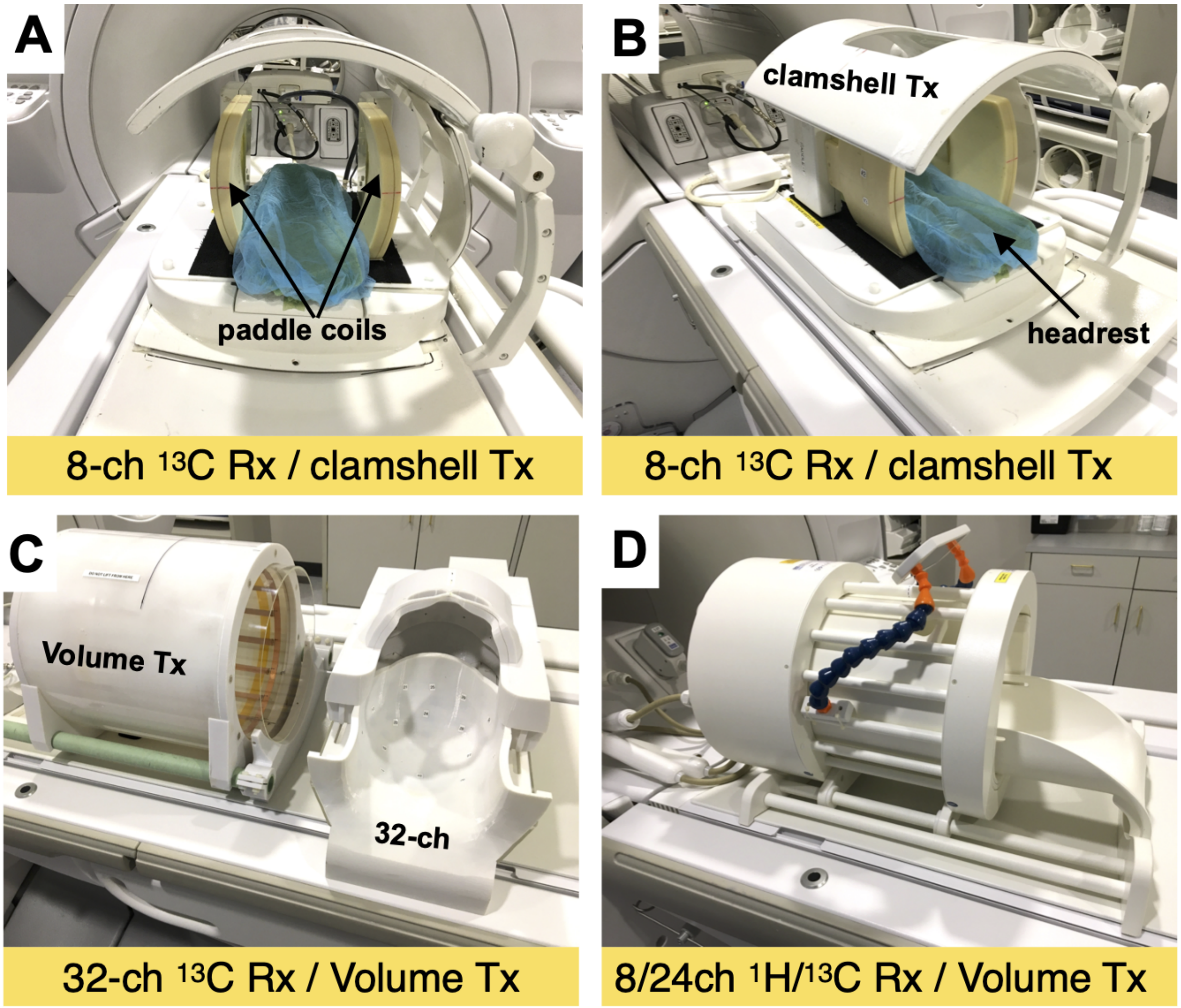

**Figure 4**: Examples of RF coil configurations used for human HP [1-$^{13}$C]pyruvate brain studies. (A,B) $^{13}$C Clamshell TX (Helmholz pair) and 2× 4-channel paddle RX arrays. (C) $^{13}$C Birdcage volume TX and 32-channel RX array (RX array slides into TX coil). (D) $^{13}$C Birdcage volume TX and 24-channel RX array, combined with a $^{1}$H 8-channel RX array. Image reproduced with permission from Ref (16).

## Phantoms

Since hyperpolarized magnetization is non-renewable, phantoms containing $^{13}$C nuclei are important to: 1) test the multi-nuclear capabilities of the imaging system, including all parts of the signal excitation and receive chain; 2) perform calibration measurements before a scan with hyperpolarized nuclei; and 3) perform necessary pre-scan adjustments (see “Prescan Calibration” section). The phantoms currently in use are listed in Table 3. Their composition must provide sufficient $^{13}$C signal, with additional considerations of conductivity, stability, chemical shift(s) present, potential for dynamic imaging, and cost. The phantom geometries are typically either compact, in order to be used alongside the subject during a HP scan, or large enough to mimic the inner volume of a RF coil for system testing.

One popular compact design contains enriched $^{13}$C-urea at high concentration, typically 8 M, which provides a single resonance, placed inside a small container ~1 mL. The most common recipe mixes $^{13}$C-urea in a 90% water/10% glycerol solution, with glycerol used to increase the urea solubility and doping with a Gd-based contrast agent to shorten $T_1$ which increases the potential SNR per unit time. For example, when Dotarem is added at a 3:1000 volume ratio the $^{13}$C-urea $T_1$ is around 500 ms and T2 is around 100 ms. However, when testing pulse sequences influenced by $T_1$ and $T_2$, doping should be used carefully. This phantom is suitable for frequency calibration, transmit gain calibration, sequence testing, and as a fiducial marker when placed next to a patient. However, enriched $^{13}$C-urea has a relatively high cost compared to natural abundance compounds.

For larger volumes (>100 ml), the phantoms most often used contain undiluted ethylene glycol, glycerol, or dimethyl silicone. These compounds have sufficiently high carbon concentrations to provide sufficient $^{13}$C signal even with the 1.1% natural abundance of $^{13}$C. These larger phantoms matching the inner volume of an RF coil are useful for coil testing, including transmit ($B_1^+$) and receive ($B_1^-$) coil profile mapping, as well as to mimic acquisitions using in vivo FOV requirements. In this case, size and conductivity should match the expected subject size in order to mimic coil loading and get a realistic estimation of $B_1$+. Large-volume natural abundance urea phantoms have also been used by some sites, but suffer from higher conductivity compared to biological tissues. Typically, it is easier to increase the conductivity and hence coil loading of the non-conductive phantom by adding NaCl to match physiological loading (16,78).

Dynamic phantoms that aim to mimic metabolite kinetics have also been developed (79–81), and have the potential to more closely mimic the HP experiment, but so far these are not widely used.

# Prescan Calibration

Prior to performing an MRI acquisition, the so-called prescan procedure is used to set the shim parameters to maximize $B_0$ homogeneity over the field of view (FOV) or a specific region of interest (ROI), the scanner center frequency (CF), the RF transmit gain, and the receiver gain. While this calibration procedure is usually automated for $^{1}$H, the lack of sufficient natural abundance $^{13}$C signal prevents use of automated methods. (Although natural abundance $^{13}$C lipid signal has been detected, there are so far no reports on using this signal for prescan.) Table 3 shows current practices across sites.

Maximizing $B_0$ homogeneity is independent of the nucleus and is therefore performed prior to $^{13}$C imaging using the $^{1}$H water signal and existing shimming tools, such as by a standard automated process (“Auto Shimming”) or using high order shimming routines. Similarly, the $^{13}$C

CF can be calculated from the $^{1}$H CF using a predetermined scaling factor that depends on the target chemical shift (82). Another common approach used is to have a small, high-concentration $^{13}$C phantom, e.g. 8M $^{13}$C-urea, integrated in the RF coil or placed next to the scan subject (1). The reference frequency can also be based on real-time measurements after the HP injection but prior to imaging (83). Both the CF and $B_0$ shimming are critical when using spectrally-selective RF pulses, as inmetabolite-specific imaging methods, where the desired excitation bandwidths are typically very narrow and frequency offsets can lead to a failure mode that is only apparent after injection.

The calibration of the RF transmit power is typically performed on a small, high-concentration $^{13}$C phantom placed near the region of interest during the scan or on a large $^{13}$C phantom of similar size and coil loading as the subject, prior to the subject scan. Reference power is often done by sweeping the power in a pulse-acquire sequence (53,62), or the Bloch-Siegert method (52,84). When using a small phantom, the location of the phantom, $B_1^+$ inhomogeneity as well as any shielding effects, e.g., when the phantom is integrated into a coil (1), may degrade the accuracy. Other methods include real-time Bloch-Siegert method measurements after the HP injection (83), and using the stronger natural abundance $^{23}$Na signal that is close enough to the $^{13}$C resonance frequency to be detected by $^{13}$C coils (82).

The receiver gain is predetermined, either systematically based on independent phantom measurements and assuming the dose and polarization of the HP compound is known prior to injection, or based on past HP imaging studies.

| Site | Primary Imaging System | $^{13}$C Receive Channels | $^{13}$C Peak Power [kW] | Phantom(s) - during study | Phantom(s) - before study | $^{13}$C Frequency Calibration | Shimming | $^{13}$C RF Transmit Power Calibration | $^{13}$C RF Receive Gain Calibration |
|---|---|---|---|---|---|---|---|---|---|
| 1 | Philips 1.5T Achieva and 3T Ingenia | 8 | 4 | $^{13}$C-urea, $^{13}$C-pyruvate, $^{13}$C-acetate, $^{13}$C-acetate, tubes | $^{13}$C-urea, tube | Derived from $^{1}$H | Auto shimming | Power sweep on phantom before study | Not required |
| 2 | GE 3T Discovery MR750w | 8 | 8 | None | $^{13}$C-urea doped with Gd, tube | Derived from $^{1}$H | Auto shimming | Bloch-Siegert | None |
| 3 | GE 3T Discovery MR750 | 8 | 8 | $^{13}$C-urea doped with Gd, sphere | $^{13}$C-urea doped with Gd, sphere | Predetermined frequency | Auto shimming | Power sweep | Empirical |
| 4 | GE 3T Discovery MR750w | 8 | 8 | [1-$^{13}$C]lactate doped with Gd<br>None | $^{13}$C-bicarbonate | Derived from $^{1}$H | PRESS box shim | Power sweep | Empirical |
| 5 | GE 3T Discovery MR750 | 32 | 8 | $^{13}$C-bicarbonate doped with Gd, sphere | ethylene glycol, head shape<br>ethylene glycol, bottle<br>dimethyl silicone, sphere & cuboid | Derived from $^{1}$H | Auto shimming<br>High order shimming | Bloch-Siegert | Empirical |
| 6 | GE 3T Discovery MR750 and | 32 | 8 | $^{13}$C-bicarbonate doped with | dimethyl silicone, various | Derived from $^{1}$H | Auto shimming | Bloch-Siegert | Empirical |

| Site | Primary Imaging System | $^{13}C$ Receive Channels | $^{13}C$ Peak Power [kW] | Phantom(s) - during study | Phantom(s) - before study | $^{13}C$ Frequency Calibration | Shimming | $^{13}C$ RF Transmit Power Calibration | $^{13}C$ RF Receive Gain Calibration |
|---|---|---|---|---|---|---|---|---|---|
| | GE 3T Signa PET/MR | | | Gd, sphere<br>$^{13}C$-urea doped with Gd, bottle or Eppendorf tube | shapes<br>$^{13}C$-bicarbonate doped with Gd, sphere<br>$^{13}C$-urea doped with Gd, bottle or Eppendorf tube | | over PRESS box (breast, prostate)<br>Auto shimming (abdomen)<br>High order shimming (brain) | | |
| 7 | GE 3T Discovery MR750w | 8 | 8 | $^{13}C$-urea doped with Gd<br>ethylene glycol | $^{13}C$-urea doped with Gd, tube | Derived from $^{1}H$ | Auto shimming followed by manual linear shimming | Power sweep | Empirical |
| 8 | GE 3T Signa Premier* | 32 | 8 | $^{13}C$-urea doped with Gd, tube | ethylene glycol | Establishing protocol | Establishing protocol | Establishing protocol | Establishing Protocol |
| 9 | GE 3T Discovery MR750 | 32 | 11 | $^{13}C$-urea doped with Gd, tube | $^{13}C$-urea doped with Gd, sphere<br>ethylene glycol, head shape<br>dimethyl silicone, sphere & cuboid | Derived from $^{1}H$ | Auto shimming<br>Specialized shimming | Power sweep | Empirical |
| 10 | Siemens 3T Biograph mMR | 12 | | $^{13}C$-urea doped with Gd, tube | $^{13}C$-urea doped with Gd, tube<br>ethylene glycol | Based on $^{13}C$ phantom | Auto shimming | Power sweep | Automatic |
| 11 | GE 3T Signa Premier | 32 | 8 | $^{13}C$-urea doped with Gd, tube | dimethyl silicone, cylinders | Based on $^{13}C$ phantom | Auto shimming | RF pulse sweep | Maximum values |
| 12 | GE 3T Discovery MR750 | 8 | 8 | $^{13}C$-urea in $D_2O$ doped with Gd | dimethyl silicone, various shapes<br>ethylene glycol, various shapes | Derived from $^{1}H$ | Auto shimming | Power sweep<br>Bloch-Siegert | Maximum values |
| 13 | GE 3T Discovery MR750w | 8 | 8 | $^{13}C$-urea doped with Gd | ethylene glycol, sphere | Based on $^{13}C$ phantom | Auto shimming | Bloch-Siegert, checked with power sweep | Maximum values |

**Table 3:** Summary of the imaging systems, phantoms, and prescan procedures used at sites currently performing HP $^{13}C$-pyruvate human studies. These were obtained from a survey of all sites performing clinical trials with HP [1-$^{13}C$]pyruvate. *Previously performed studies with a Siemens 3T Tim Trio. The imaging systems, phantoms, and prescan procedures reported in the reviewed papers are shown in Supporting Table S1.

## Summary

Commercially available 3T MRI systems are by far the most commonly used for human HP $^{13}$C-pyruvate studies, although a systematic investigation of the impact of $B_0$ has only recently been investigated (73). The multi-nuclear RF transmit and receive chain has proven sufficient for current acquisition strategies, although many sites have observed artifacts due to RF interference, gradient interference, and residual eddy currents when operating at the $^{13}$C frequency. A variety of $^{13}$C RF coils, tailored for numerous anatomical targets, have been successfully demonstrated, with the main limitation that most transmit coils take up a lot of additional space inside the bore and provide relatively inhomogeneous $B_1^+$ profiles. The phantoms used have converged into generally 2 categories - small phantoms containing $^{13}$C-enriched compounds that can be used during the study and human-sized phantoms containing compounds with high carbon concentrations but without $^{13}$C enrichment that are used to test and calibrate the coils. There are no standardized compositions or geometry, and dynamic phantoms that recapitulate in vivo kinetics would be desirable but are still an emerging area. Prescan calibration procedures were not well defined in most publications, so we surveyed individual sites to determine current practices. Calibration procedures for the $B_0$ field ($^{13}$C CF and shimming) for most sites take advantage of $^1$H signal and methods, while methods for calibration of $B_1^+$ is more variable across sites, likely a reflection of remaining challenges in how to perform this calibration. Standardization of both phantoms and calibration procedures would synergistically improve the robustness and reproducibility of HP $^{13}$C studies.

# Acquisition and Reconstruction

Data acquisition strategies in human HP [1-$^{13}$C]pyruvate MRI studies must account for multiple chemical shifts, efficiently utilize the non-renewable HP magnetization, and acquire data quickly relative to metabolism and relaxation decay processes. These studies require spectral encoding to separate metabolites, necessitating pulse sequences that efficiently encode up to 5D data (3 spatial + 1 spectral + 1 temporal dimension). RF pulses must efficiently sample without immediately saturating the non-renewable HP magnetization, and sequences must acquire data quickly and be robust to both experimental and physiologic variation (e.g. $B_1^+$ inhomogeneity, variation in perfusion) to ensure reproducibility and minimize scan-to-scan variability. This section covers current successful practices for data acquisition in human [1-$^{13}$C]pyruvate studies, and accompanying $^1$H imaging, from different anatomic regions, including scan parameters and image reconstruction.

## Acquisition and Reconstruction Methods

The acquisition methods used in human [1-$^{13}$C]pyruvate studies can be classified into 3 categories: 1) MR spectroscopy or MR spectroscopic imaging (“MRS/I”), 2) chemical shift encoding methods, and 3) metabolite-specific imaging (Fig. 5). MRS/I methods specifically resolve a spectrum that can be analyzed to extract expected as well as unexpected resonances, making this approach very robust. It was used in many initial studies (1). Chemical shift encoding methods, most commonly the Iterative Decomposition of water and fat with Echo

Asymmetry and Least-squares estimation (IDEAL) method, use imaging sequences acquired with multiple TEs and rely on a model-based separation of expected chemical shifts (85). Metabolite-specific imaging methods use specialized RF pulses that are spatially and spectrally selective to excite individual metabolites which are then typically imaged with fast k-space trajectories such as echo planar imaging (EPI) or spirals (86). Their application to different organ systems is described below.

The image reconstruction methods used in human [1-$^{13}$C]pyruvate studies have typically been conventional methods (e.g. FFT, non-uniform FFT, or equivalent). The incorporation of accelerated imaging and advanced reconstruction methods including parallel imaging (4,57,87) and compressed sensing (7) has also been applied in human studies for improved spatial resolution, temporal resolution and coverage, but have the potential for additional artifacts as well as SNR losses due to ill-conditioning of the reconstruction (e.g. g-factor). The majority of published studies do not use accelerated imaging indicating the resolution and coverage achievable without acceleration is currently adequate for successful data collection. Performing coil combination, even with fully sampled data has also been shown to have specific challenges for HP human images: using naive sum-of-squares methods suffer from high noise amplification in the relatively low SNR regime of HP [1-$^{13}$C]pyruvate (compared to $^{1}$H), motivating several HP $^{13}$C-specific methods that include data-driven coil sensitivity estimation which have shown obvious improvements over sum-of-squares (11).

More recently denoising techniques have been applied as post-processing of human HP data(41,42,44). The techniques applied are based on spatial-temporal singular value decomposition for unsupervised estimation of signal and noise components. They have shown improvements in apparent SNR in the brain and liver, while care must be taken to choose parameters such as the rank threshold to avoid oversmoothing and overfitting to the estimated signal components.

# Prostate Studies

Prostate cancer was the first human application of HP [1-$^{13}$C]pyruvate (1), and data was acquired with MRS/I methods: 1D dynamic MRS, single-slice 2D dynamic echo-planar spectroscopic imaging (EPSI), and single time point 3D EPSI. Advances in imaging strategies led to the development and application of new acquisition schemes, including undersampled 3D EPSI with compressed-sensing (7), model-based chemical shift encoding methods that use *a priori* information (47,59), and metabolite-specific EPI (10), all of which can provide volumetric whole-organ coverage and dynamic acquisitions.

The pyruvate bolus arrival in the prostate can vary by ± 10 s between patients, necessitating dynamic imaging to reliably and consistently capture the pyruvate bolus (18). For this reason, all currently ongoing studies acquire dynamic data. While MRS/I, chemical shift encoding, and metabolite-specific imaging can all achieve dynamic imaging, chemical shift encoding and metabolite-specific imaging provide greater dynamic and volumetric coverage (85). For scan prescriptions, the FOV is designed to provide full prostate coverage and typically to match the orientation of the anatomic imaging used for registration. Flip angles used in current studies are constant through time, as quantification with a variable-through-time flip scheme is highly sensitive to bolus timing (8) and errors in the RF transmit ($B_1^+$) field (76).

## Heart Studies

Data acquisition methods for $^{13}C$ imaging in the heart must be designed to meet the demands of significant cardiac motion and blood flow. To cope with the periodic cardiac motion, most human heart studies to date used gating to the diastolic window, the longest cardiac cycle interval, which has reduced motion (2,22,28,30,35,36,38,45,52). The duration of the diastolic window limits the available data sampling time, making cardiac acquisitions the most time-constrained of the HP $^{13}C$ MRI applications. The most common acquisition approach is metabolite-specific imaging with spiral k-space trajectories (2). Their single-shot imaging capability makes these methods particularly robust to motion effects. Furthermore, spiral k-space trajectories provide rapid k-space coverage and relatively benign flow and motion artifacts. The majority of studies have used 2D multi-slice acquisitions, but 3D encoding has also been used successfully (35).

## Brain Studies

For HP $^{13}C$ MRI of the human brain, the majority of studies have also used 2D (slice selective) acquisitions (10–12,14,16,28,33,40,41,44,51,53,60), with a trend toward volumetric coverage using 2D multi-slice metabolite-specific imaging. 3D metabolite-specific imaging of the whole brain, with phase encoding of the slice direction (34,57), has been shown to provide similar SNR efficiency (88) compared with multislice imaging. A number of studies have employed MRS/I (5,6,29,31–33,50,55) resulting in a spectrum from each voxel, which has the advantage of not requiring a priori information about which peaks to encode. This was important in early brain studies when it was not known which peaks would be detectable. Chemical shift encoding, using a set of images with different echo times and an iterative reconstruction of the individual resonances (i.e. the IDEAL approach (85)), has also been used (12,49,54), with the drawback that coverage in the slice direction was limited due to the time required to acquire multiple echo time images.

## Abdomen and Breast Studies

The fundamental approaches to data acquisition and reconstruction in the abdomen and breast are largely similar to the aforementioned applications, but demand attention to particular challenges associated with these anatomic regions, especially relating to respiratory motion.

Although it has been shown that a basic 2D MRSI approach based on phase encoding and FID readout can be successfully applied for HP $^{13}C$ imaging in breast (15) and kidney (13), major advantages in terms of spatiotemporal resolution and coverage have been realized using tailored approaches based on metabolite-specific imaging (43,62) and chemical shift encoding (43), which have facilitated multi-slice or 3D dynamic acquisitions over large FOVs in the abdomen (4,37,46).

The significant respiratory motion encountered in these regions can directly blur $^{13}C$ images, and has further favored these rapid acquisition strategies. Motion also degrades $B_0$ homogeneity, which can shift frequency-selective excitation profiles and introduce artifacts into rapid imaging readouts. This makes accurate determination of the acquisition center frequency and shimming essential in these regions which often cover large FOVs. (See “Prescan Calibration” section for more information). In some studies, breath-holding was used to minimize motion effects and enforce frame-to-frame data consistency (42). A pragmatic and reasonably effective approach for dealing with respiratory motion during $^{13}C$ data acquisition is an initial breath-hold (as long as can be tolerated), followed by free-breathing (46,62).

## $^{1}H$ Imaging

Collection of $^{1}H$ imaging data is essential both for prescribing the $^{13}C$ acquisition and for interpretation of the resulting $^{13}C$ data. Multi-planar $^{1}H$ scouts are acquired prior to $^{13}C$ acquisition to enable graphical prescription of the $^{13}C$ imaging region. All human HP $^{13}C$-pyruvate imaging studies acquire conventional MRI scans (e.g. $T_1$- and $T_2$-weighted volumes) for anatomic reference, aiming to cover at least the full $^{13}C$ FOV. Acquiring these anatomic scans as close as possible to the time of $^{13}C$ imaging (immediately before or after) minimizes potential misregistration between the data sets. Depending on the application, other advanced $^{1}H$ sequences are also acquired (e.g. diffusion-weighted imaging for cancer imaging). When contrast-enhanced data is acquired, it is done after $^{13}C$ imaging, as paramagnetic contrast agents will accelerate $^{13}C$ relaxation.

## Reported Study Parameters

Figures 5 and 6, and Supporting Table S2 shows the reported acquisition study parameters for human HP [1-$^{13}C$]pyruvate studies published as of September 2022. Figure 5 shows a mixture of MRS/I, metabolite-specific imaging, and chemical shift encoding methods have been successfully used, where spectroscopy-based methods have become less prevalent in recent studies. Figure 6 shows the acquisition timing, including the important start time and interval/temporal resolution, is quite variable across studies.

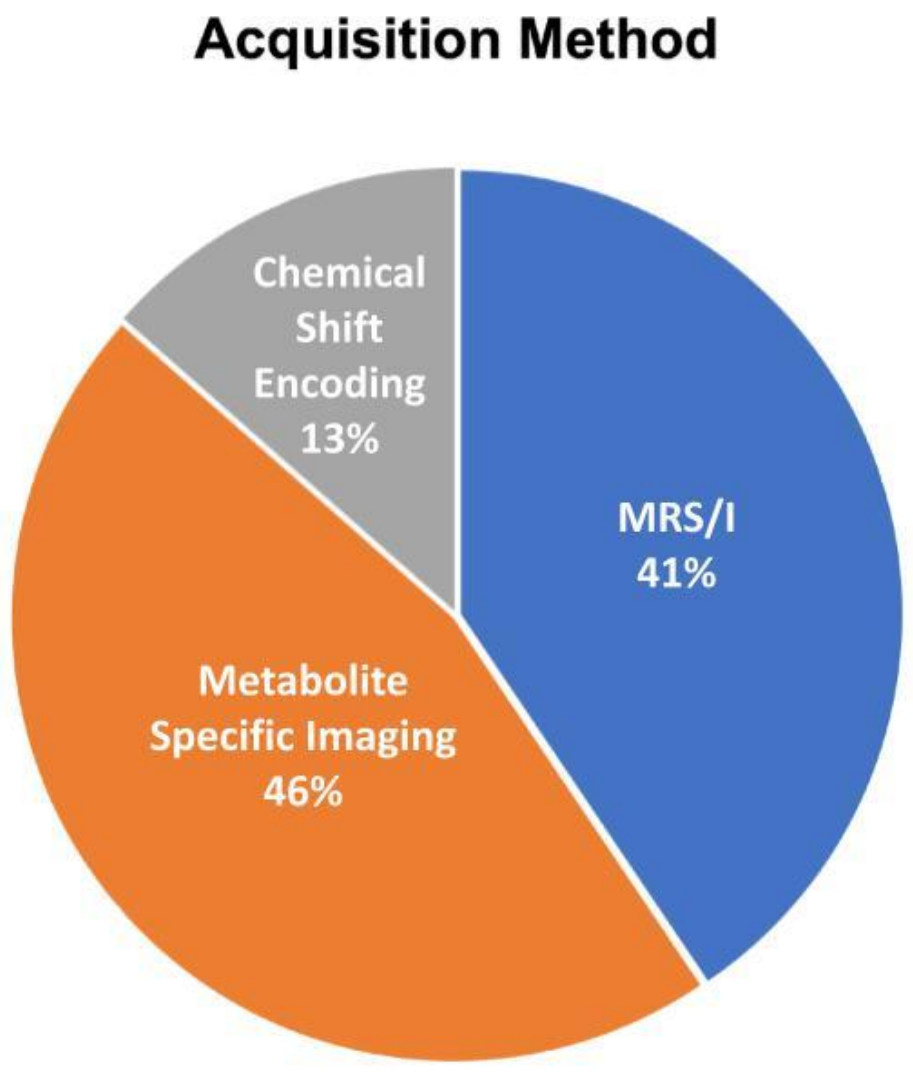

**Figure 5**: Acquisition methods used in published HP [1-$^{13}$C]pyruvate human studies published up to September 2022, classified into: MR spectroscopy and spectroscopy imaging (MRS/I); chemical shift encoding methods, such as IDEAL, that use multiple TEs and model-based reconstructions; and metabolite-specific imaging methods that use spectrally-selective excitation to image a single resonance at a time.

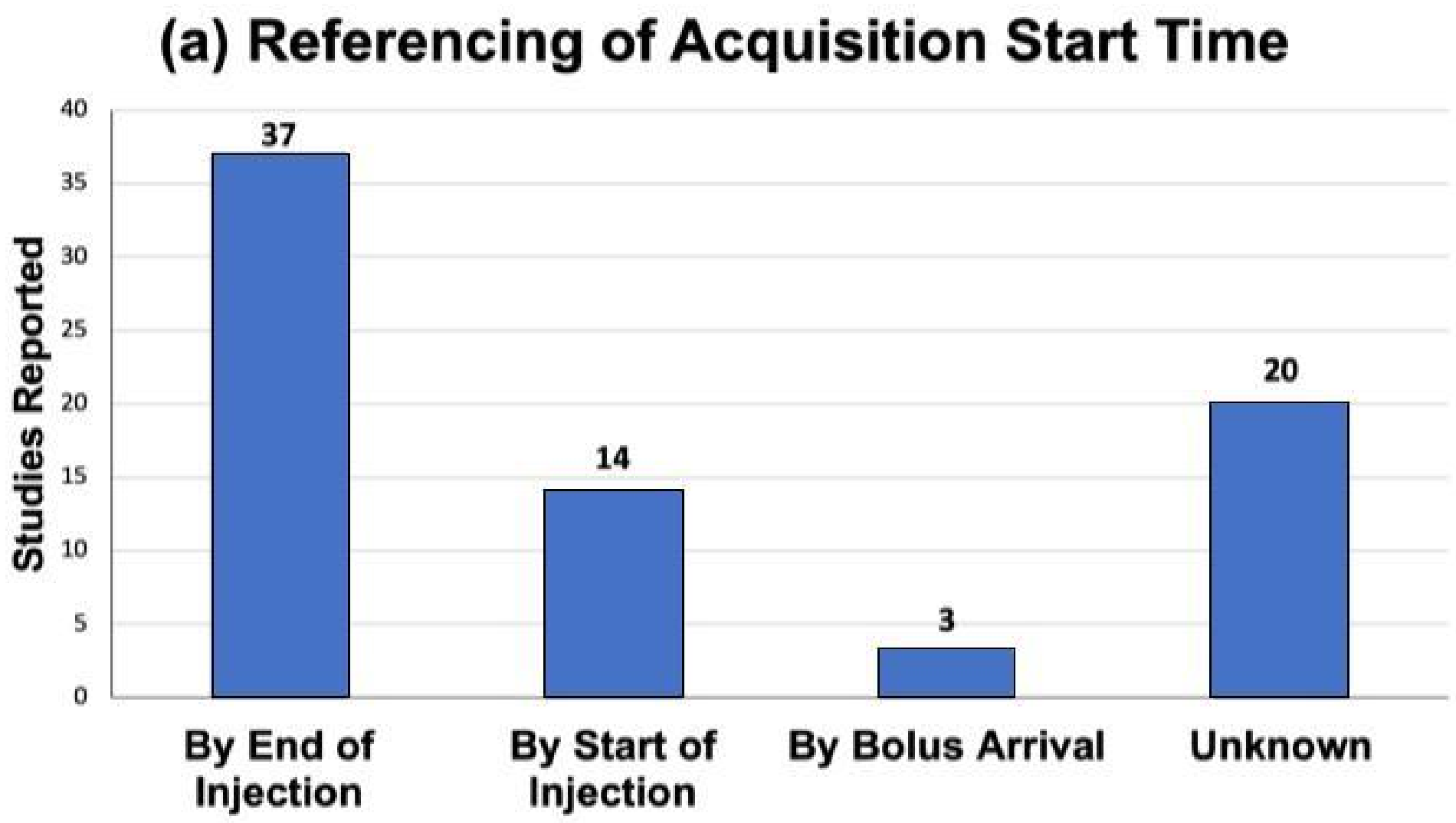

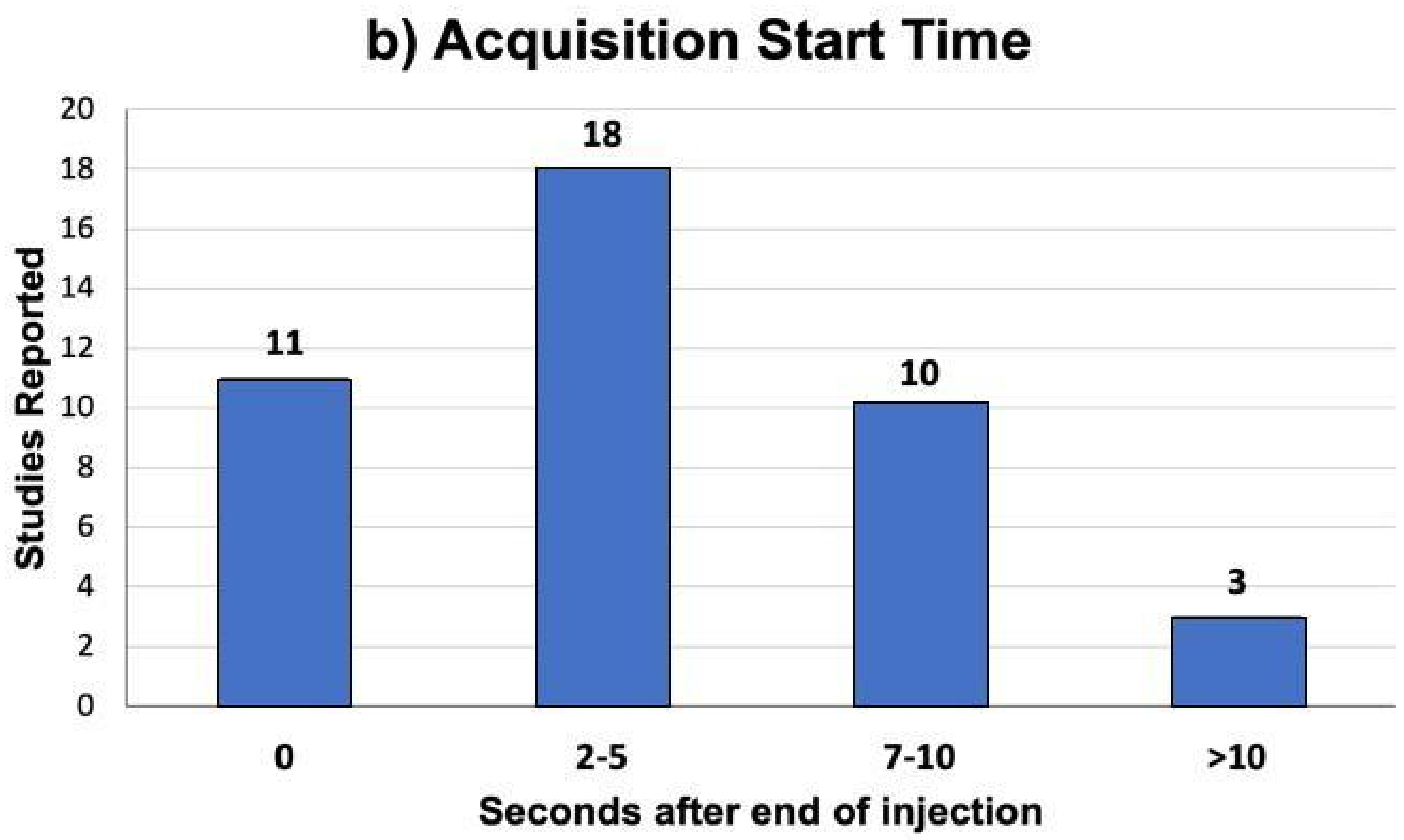

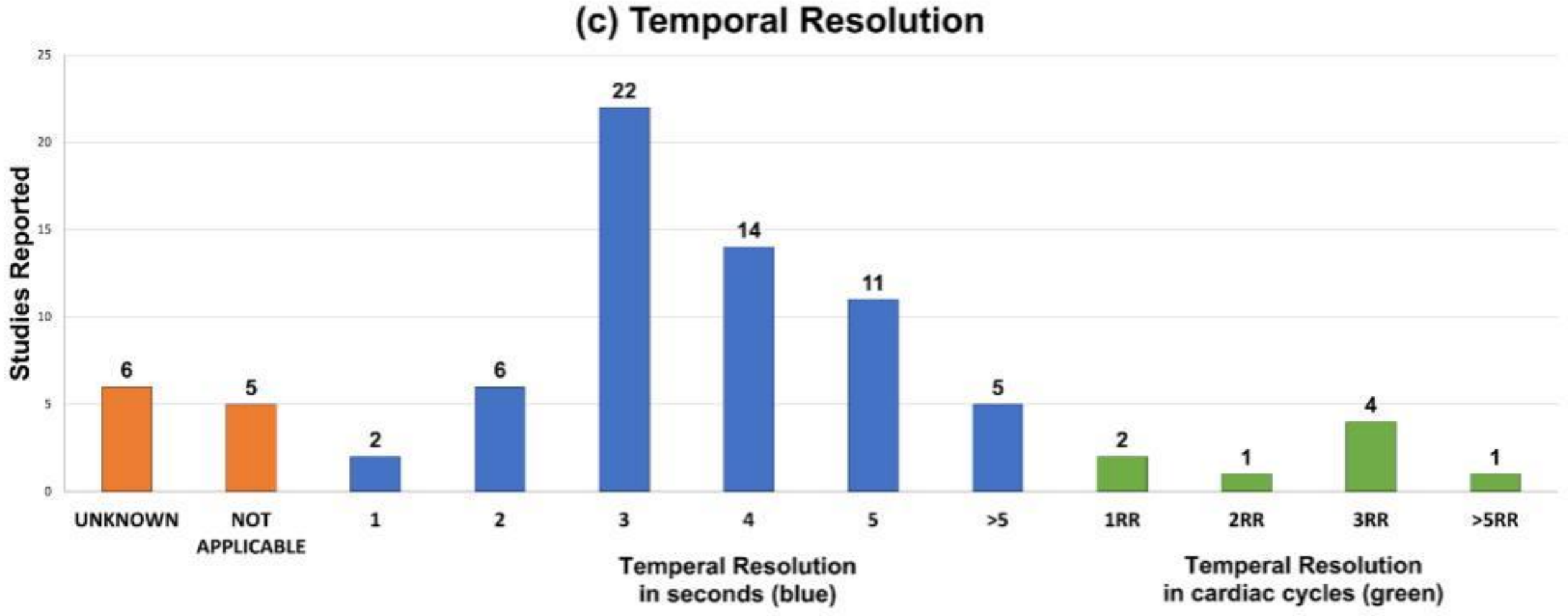

**Figure 6**: Temporal acquisition characteristics reported in HP [1-$^{13}$C]pyruvate human studies published up to September 2022. (a) Reported referencing of acquisition start times. (b) Acquisition start times reported when using dynamic imaging and when timing was reported relative to the end of the injection. (c) Temporal resolutions. "Not Applicable" indicates dynamic imaging was not used.

# Summary

Three general categories of acquisition strategies have been used successfully for human HP $^{13}$C-pyruvate studies: MRS/I, model-based chemical shift encoding (e.g. IDEAL) methods, and metabolite-specific imaging methods. These have enabled successful studies in the prostate, heart, brain, abdomen, and breast. Recent studies increasingly have used the imaging-based strategies of metabolite-specific imaging and chemical shift encoding which are the fastest methods, although a heads-to–head comparison between techniques has not been performed. Metabolite-specific imaging is quite popular because of its speed and compatibility with single-shot imaging, but is sensitive to $B_0$ field variations and thus requires careful calibrations. Nearly all studies surveyed acquired data dynamically, allowing measurement of the bolus and metabolite kinetics. The exact timings and associated flip angles vary quite widely across reported studies, with no consensus yet as to how to choose these parameters. Image reconstruction is typically done directly using Fourier Transform methods, and accelerated imaging strategies are uncommon.

# Data Analysis and Quantification

This section covers the analysis of data from human HP [1-$^{13}$C]pyruvate studies, including modeling and metrics, visualization, as well as considerations for how to store data and metadata. Depending on study design, the analysis may need to give quantitative or semi-quantitative output reflecting a biological process or may just reflect a contrast between different regions of interest for quantitative evaluation.

## Metrics

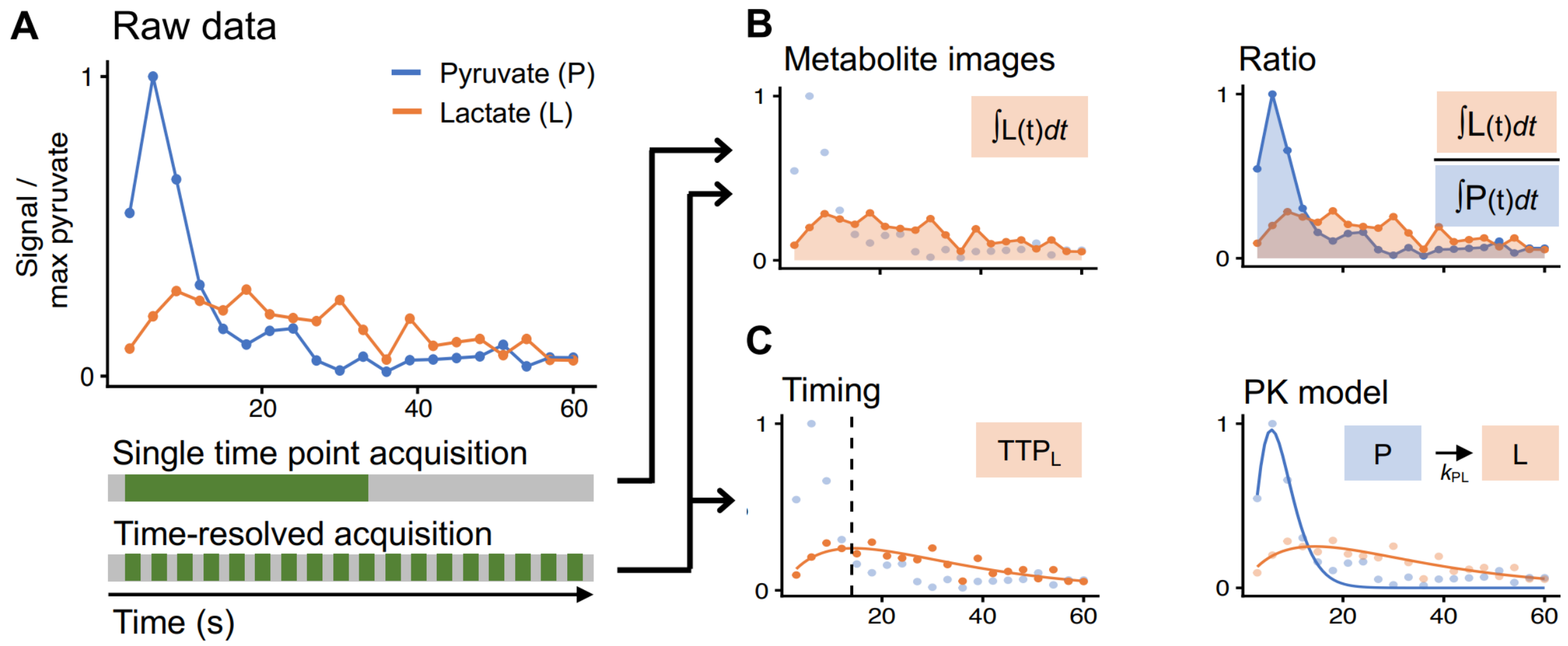

**Figure 7**: HP [1-$^{13}$C]pyruvate raw data (A) have typically been quantified using four categories of metrics depending on the acquisition. Data acquired as a single time point are often quantified using normalized metabolite images or metabolite ratios (B). Dynamic data can be quantified using normalized metabolite images or metabolite ratios (B), or with metabolite timings such as time-to-peak (TTP) or pharmacokinetic (PK) models (C). The latter two require

the data to be time-resolved. [1-$^{13}$C]alanine and $^{13}$C-bicarbonate are analyzed similarly to [1-$^{13}$C]lactate but omitted here for display.

Metabolite images are commonly used as summary metrics for HP MRI data, often including some form of normalization as well as summed over time as an area under the time curve (AUC) (17). These are analogous to the visual evaluation that is most used for routine clinical work (89,90). In these metabolite images, we expect that the [1-$^{13}$C]pyruvate AUC signal is predominantly weighted towards perfusion and uptake, while [1-$^{13}$C]lactate, [1-$^{13}$C]alanine and $^{13}$C-bicarbonate AUCs represent metabolic conversion. The strength of this approach lies in its simplicity and relatively few underlying assumptions. Limitations to the use of single-metabolite images or AUCs include sensitivity to inhomogeneous coil profiles (57,87,91), the acquisition strategy and acquisition parameters, pyruvate polarization and concentration level, and signal relaxation rates (92). Further, the reader must be careful to interpret all the images in conjunction to better understand the underlying biology; for example, increased [1-$^{13}$C]lactate in the presence of decreased [1-$^{13}$C]pyruvate delivery can have a very different meaning compared to increased [1-$^{13}$C]lactate with increased [1-$^{13}$C]pyruvate delivery.

In an attempt to address variations in coil sensitivity, polarization level, and pyruvate delivery, AUC images are often computed by normalizing to a specified parameter, such as the maximum pyruvate or average lactate signals, or presented as a ratio such as lactate/pyruvate or divided by “total Carbon” - the sum total of HP $^{13}$C signal observed across all metabolites. The AUC ratios between metabolites and pyruvate are proportional to the corresponding forward kinetic rates (81,93), but are not directly comparable to rate constants when magnetization loss rates (e.g. relaxation and losses due to signal excitation) differ between studies. Similarly, the ratios between the produced metabolites (e.g. bicarbonate/lactate) can reflect the balance between downstream metabolic pathways (12,55). Care must be taken to consider how AUC images are calculated and normalized before comparing values between studies.

To further quantify the interpretation, pharmacokinetic (PK) modeling approaches were developed to compute the apparent kinetics of pyruvate-to-metabolite exchange (92,94–99). These yield semi-quantitative to quantitative apparent rate constants, given in $s^{-1}$. Some models require a vascular input function, while others avoid this requirement (95). PK models can explicitly account for acquisition-specific details such as excitation angle and repetition time, and thus may reduce the effects of these details on quantification. An input-less model, provided in the Hyperpolarized-MRI-Toolbox (https://github.com/LarsonLab/hyperpolarized-mri-toolbox) (100) and thus frequently employed for human data, has been shown to fit well and robustly to prostate and brain data (8,20). PK models are quantitative in nature, arguably provide more relevant biological information (8,20), and appear to be reproducible across sites (51). However, rate constants derived from PK models are still apparent rates, and likely do not reflect a single biological characteristic.

Some additional considerations include whether complex or magnitude data is used, as the noise behaviors will impact the analysis differently. Additionally, cut-off thresholds or other criteria may be used to identify and avoid voxels with insufficient SNR before analysis to improve robustness (20,41).

Regardless of the analysis approach, the underlying biology is not always clearly represented by the data; instead, the metrics may be influenced by perfusion, barrier permeability, intercellular shuttles, enzyme activities, co-substrate concentrations, or combinations thereof, depending on the organ and disease of interest (19,43,94,101–103). This may be addressed by incorporating complementary information. As an example, HP $^{13}$C pyruvate data is influenced by perfusion, and thus addition of perfusion MRI could be important for interpretation (98,104,105).

All the methods outlined above have been explored in clinical studies, described in Supporting Table 3 and summarized in Figure 8. As of September 2022, approximately 52% of studies involving human subjects report rate constants derived from a PK model with a few different models reported. A nearly equal fraction (51%) of the studies report AUC ratio values. Approximately 66% of these studies report metabolite-specific images or AUC values. About 40% report SNR values; this metric is particularly frequent in manuscripts that describe technical developments for clinical HP MRI. Approximately 16% of these studies summarize model-free metrics, and 10% report measurements from a single timepoint. Most studies report a combination of quantities.

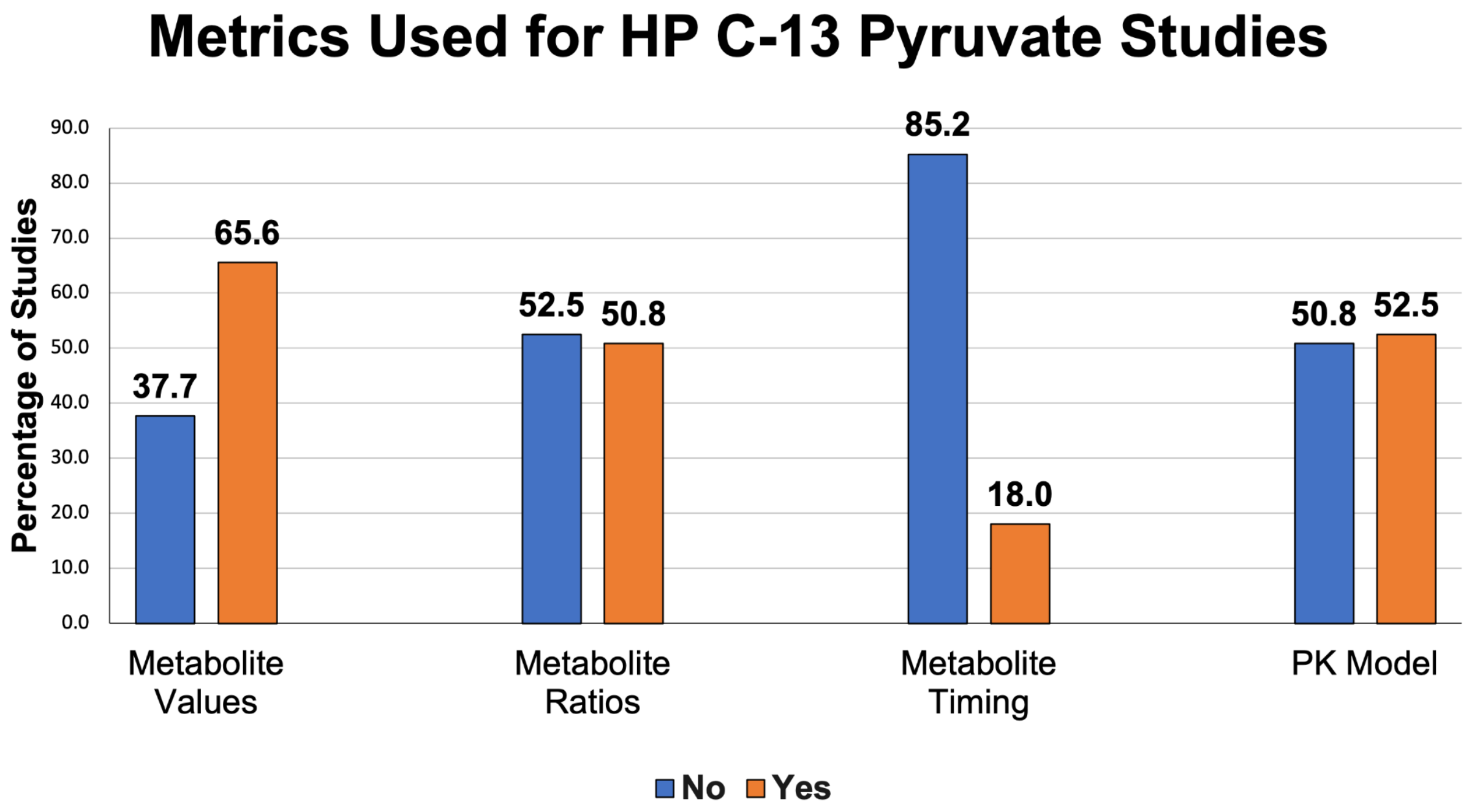

**Figure 8**: Reported metrics used for analysis in HP [1-$^{13}$C]pyruvate human studies published up to September 2022.

# Visualization

A wide variety of approaches have been used for visualizing data from human HP $^{13}$C-MRI studies. The challenges and practical considerations are: 1) choosing the appropriate metrics to display, 2) how to encode the parameters (e.g. the colormap), and 3) choosing how to provide anatomical context and other multi-parametric data. The choice of visualization also depends on the goal which could be for diagnostic interpretation, but also quality control, reproducibility among readers and publication.

*Metrics*

The choice of HP $^{13}$C metrics is described in detail above. At this stage in HP 13C development where there is no standardized metric, often a combination of metabolite images and ratios or PK model parameters are shown.

*Parameter Encoding*

The mapping function chosen should provide an adequate, often quantitative, impression of the parameter mapped. There is a consensus in the visualization field that perceptually uniform maps are best suited to visualize continuous parameters, like the greyscale typically used by radiologists as well as other monochrome (black to blue) and color ranges (fire-type, rainbow-type) (106,107). Multi-color heatmaps have been the most frequently employed method for HP $^{13}$C data, while greyscale has infrequently been used but it ensures there is no coloring-based bias as well as facilitating later reuse (Fig. 9a). Among the color schemes employed in the clinical HP $^{13}$C literature, fire-type scheme seems to be the most common [similar to “Plasma” or “Inferno” in matplotlib.org]. Next most commonly employed is the rainbow-type scheme [similar to “Rainbow” in matplotlib.org].

*Anatomical Context*

HP MRI faces the challenge that it does not necessarily depict the anatomical features, similar to PET, and thus requires an anatomical reference. Most often, a grayscale anatomical image is overlaid with a HP colormap (Fig. 9c,d). This approach is very intuitive, but can skew perception as the grey-scale anatomical reference may affect the brightness of the HP data (e.g. signal in the skull). This bias does not occur when showing adjacent maps (Fig. 9a, b). Here, anatomical outlines may help to provide reference (Fig. 9b).

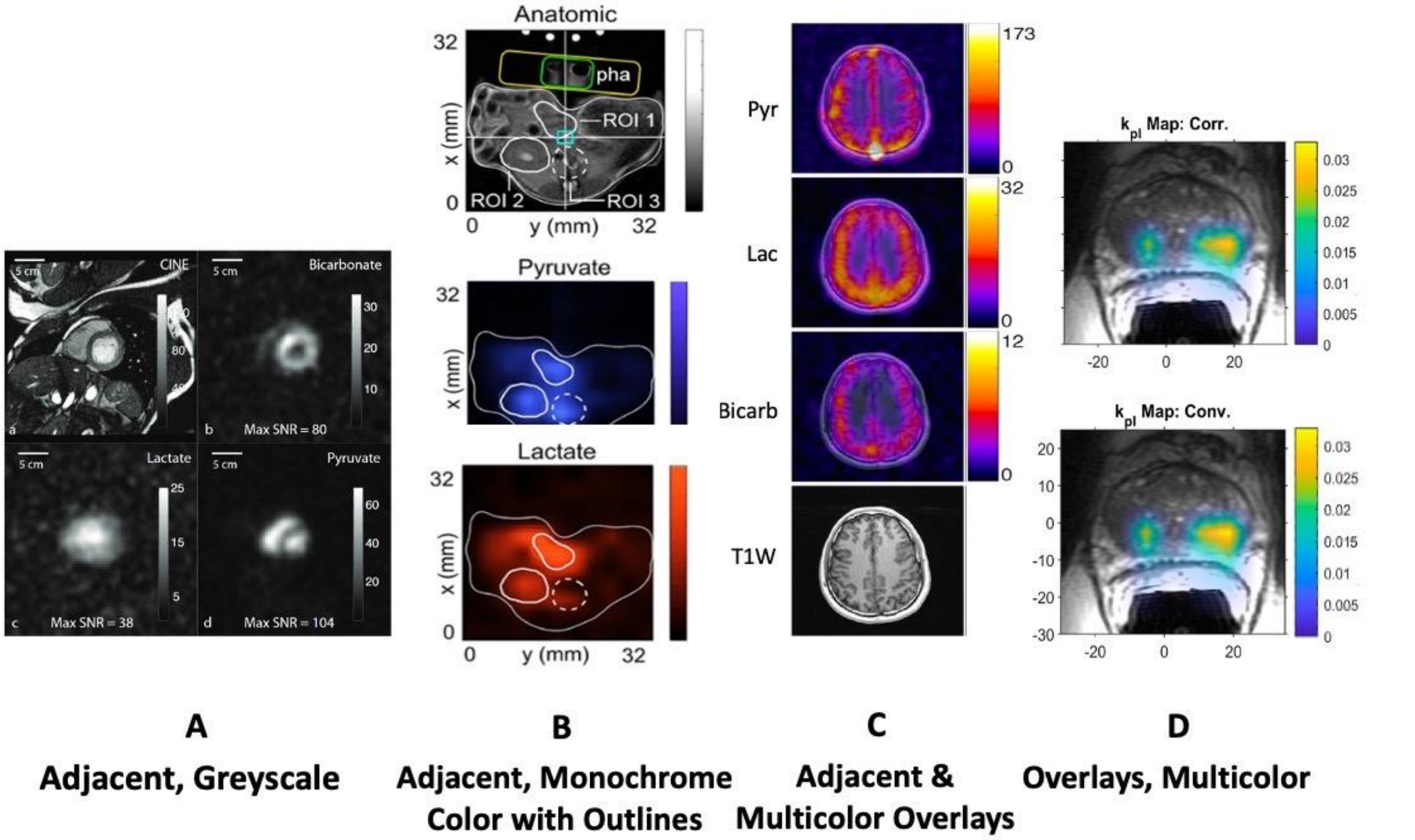

**Figure 9**: Examples of different parameters, color schemes and anatomical reference used to visualize HP $^{13}$C MRI: (A) cardiac study showing each metabolite (SNR) on one monochromatic grayscale image and separate anatomic reference, no overlays (2); (B) preclinical cardiac study showing one image per metabolite in monochrome (red / blue) on outlines of anatomy, and additional annotated anatomical image in separate panel (108); (C) brain study where each metabolite AUC (multicolor) was fused with an anatomic image, plus one anatomical image (28); (D) prostate study showing $k_{PL}$ as multicolor overlay on anatomic image (26).

# Summary

Three general categories of analysis metrics have been reported: metabolite images (often normalized), metabolite ratios, and PK model parameters. There is no apparent convergence to one of the categories, and many studies report multiple metrics in their analysis. A standard metric that is robust to experimental variations (known or unknown) would be highly desirable to compare results across studies. PK modeling of conversion rates ($k_{PL}$, $k_{PB}$) have shown the most promise in this regard. HP $^{13}$C-pyruvate data has always been visualized with $^1$H anatomical reference images, allowing for localization of metabolism information. Most papers show HP data as color overlays, but using a variety of colormaps. Using a consistent, perceptually-uniform colormap would facilitate better translation of HP $^{13}$C-pyruvate MRI, and we can probably learn from clinical PET practice where metabolism data is also visualized alongside anatomical reference images.

# Discussion and Conclusions

Hyperpolarized [1-$^{13}$C]pyruvate MRI has made great strides in recent years, with an explosion of human research studies covering a broader range of applications and performed by more sites worldwide. We have systematically surveyed the methods for HP agent preparation, system setup and calibration, acquisition and reconstruction, and data analysis used in previous and current studies. The results of this in this paper highlights that the areas of common procedures include the use of the same clinical hyperpolarizer, $^{13}$C phantoms configurations for testing and calibration, time-resolved imaging with a trend towards imaging-based acquisitions, and the use of area-under-curve and PK model based metrics. It also highlights that the areas of variations include the fluid path preparation approach and agent release criteria, system and experiment calibration procedures, choice of imaging-based acquisition and acquisition timing, and numerous metrics applied for analysis. See the summary for each section above for additional details.

In preparing this paper we also observed notable omissions in reporting of study methods that we believe are relevant to the study results. It was often difficult to determine the calibration procedures used. HP pyruvate preparation processes were often minimally described or left out entirely. In some studies, the key parameter of acquisition timings was not reported. Based on these findings, we believe it would be valuable to define guidelines for reporting of studies to ensure a minimum set of methods is described.

There is some potential bias in our findings and reported methods because one site (UCSF) has published more human HP [1-$^{13}$C]pyruvate than other sites, accounting for 25 out of 63 of the papers reviewed. To mitigate this bias, we show the results of current practices submitted from different sites and have included named authors representing a variety of sites as well.

The next steps we propose are formal consensus building, in which a voting process will be used in the hopes of creating a concrete set of recommendations and best practices, as well as clearly identify where there is a lack of consensus and need for further development. It's important to note that HP $^{13}$C-pyruvate MRI is still a research study, and therefore, the methods described and recommended will be treated as research from a regulatory point of view. With the development of guidelines and best practices, we hope to improve the reliability and reproducibility of results in future studies, and facilitate the translation of this technology to clinical applications.

# Acknowledgements

In addition to the named authors, this work was supported and reviewed by the HP 13C MRI Consensus Group. This group includes the named authors as well as:
Aarhus University, Aarhus, Denmark: Esben Søvsø Szocska Hansen, Jack Miller, Lotte Bonde Bertelsen, Michael Vaeggemose
Chang Gung Memorial Hospital, Taojuan City, Taiwan: Gigin Lin

Technical University of Denmark, Copenhagen, Denmark: Andrea Capozzi
ETH Zurich, Zurich, Switzerland: Max Fuetterer
GE Healthcare, Waukesha, Wisconsin, USA: Albert Chen, Avi Leftin, Rolf Schulte
Institute for Bioengineering of Catalonia, Barcelona, Spain: Irene Marco Rius
Kings College London, London, UK: Thomas C Booth
École Polytechnique Fédérale de Lausanne, Lausanne, Switzerland: Mor Mishkovsky
MD Anderson Cancer Center, Houston, Texas, USA: Christopher Michael Walker
Memorial Sloan Kettering Cancer Center, New York, New York, USA: Kayvan Keshari
MIAAI (Medical Image Analysis and AI), Danube Private University, Kerns-Stein an der Donau, Austria : Ramona Woitek
National Institutes for Quantum Science and Technology, Chiba, Japan: Yuhei Takado
National Institutes Health/National Cancer Institute, Bethesda, Maryland, USA: Crystal Vasquez
NVision imaging technologies, Ulm, Germany: Christoph Müller
Oxford Instruments, Oxford, UK: Joel Floyd
Technical University of Munich, Munich, Germany: Martin Grashei, Geoff Topping, Frits van Heijster
University of California - San Francisco, San Francisco, California, USA: Jenny Che, Yaewon Kim, John Kurhanewicz, Michael Ohliger, Renuka Sriram, Myriam Chaumeil
University of Freiburg, Freiburg, Germany: Michael Bock, Andreas Schmidt
University of Cambridge, Cambridge, UK: Ferdia Gallagher
University of College London, London, UK: Rafat Chowdhury, Yangcan Gong, Shonit Punwani
University of Florida, Gainesville, Florida, USA: Matthew Merritt
University of Oxford, Oxford, UK: James Grist, Damian Tyler
University of Texas - Southwestern, Dallas, Texas, USA: Fatemeh Khashami, Craig Malloy, Jae Mo Park, Vlad Zaha

This work was also supported by the ISMRM Hyperpolarized Media MR Study Group and the ISMRM Hyperpolarization Methods & Equipment Study Group.

# Supporting Tables

**Supporting Table S1**: MRI system setup and calibration methods reported in HP $^{13}$C-pyruvate human study papers surveyed. The entries are blank when the methods were not reported in the manuscript or supplementary/supporting materials.

**Supporting Table S2**: Acquisition methods reported in HP $^{13}$C-pyruvate human study papers surveyed. Note that some papers reported multiple acquisition methods. The entries are blank when the methods were not reported in the manuscript or supplementary/supporting materials. If some values were not reported, they are denoted by a "?". In some cases, the methods were ambiguous and have been inferred from the cited paper or its references, which are *italicized* and followed by "?". "~" indicates a range of parameter values were reported. N/A = Not applicable. For the Start Time, some studies reported a single value, but it was not clear whether this was relative to the start of injection, end of injection, or other time, and these are shown as a single value and are *italicized*.

**Supporting Table S3**: Metrics reported in HP $^{13}$C-pyruvate human study papers surveyed.

## Supporting Information for “Current Methods for Hyperpolarized [1-13C]pyruvate MRI Human Studies”

**Supporting Table S1**: MRI system setup and calibration methods reported in HP $^{13}$C-pyruvate human study papers surveyed. The entries are blank when the methods were not reported in the manuscript or supplementary/supporting materials.

*Abbreviations*: TX = transmit coil; RX = receive coil; TG = transmit gain; CF = center frequency

*Coil Vendors*: GE = GE Healthcare, Waukesha, WI, USA; RAPID Biomedical = RAPID Biomedical GmbH, Rimpar, Germany; Clinical MR Solutions = Clinical MR Solutions, Brookfield, WI, USA; Invivo = Invivo Inc.; PulseTeq = PulseTeq Limited, Chobham, Surrey, UK

| Ref | Imaging System | RF coils 13C | RF coils 1H | Phantoms | Pre-scan |
|---|---|---|---|---|---|
| 1 | GE 3T Discovery MR750 | TX 13C clamshell<br>RX 1H/13C endorectal coil (GE) | 1H body coil, 1H/13C endorectal coil (GE) | 8M 13C urea | signal calibration using the 8M 13C urea phantom |
| 2 | GE 3T Discovery MR750 | TX 13C volume transmit coil system (GE)<br>RX Two 4-channel 13C surface coils arrays (GE) | | spherical phantom containing an ≈8 mol/L solution of 13C-urea, which was fixed on top of the anterior receiver coil housing | prescan calibration of the 13C receive frequency and transmit power was performed using the signal from a 1.5-cm diameter spherical phantom |
| 3 | GE 3T Discovery MR750 | | | | |
| 4 | GE 3T Discovery MR750 | TX 13C clamshell (GE)<br>RX 16-channel bilateral phased array (RAPID) | | | |
| 5 | GE 3T Discovery MR750w | TX 13C clamshell (GE)<br>RX Two 4-channel 13C surface coils arrays (GE) | 8-channel 1H transmit/receive head coil | | |
| 6 | GE 3T Discovery MR750 | TX 13C clamshell (GE)<br>RX Two 4-channel 13C surface coils arrays (GE) | | unenriched ethylene glycol + sealed standard that is housed within one of the eight-channel phased array elements and contains 1 mL of 8 M of 13C-urea. | Frequency calibration and B1+ map with phantom |
| 7 | GE 3T Discovery MR750 | TX 13C clamshell<br>RX 1H/13C endorectal coil (GE) | 4-channel pelvic phased coil array + 1H-tuned endorectal coil element | 13C-urea phantom positioned on the receive coil (8 M, 600 μL) and 2 ethylene glycol phantoms (natural abundance, 13C concentration = 0.17 M) | 13C-urea reference used for 13C RF calibration and center frequency, 1H MRSI shimming |
| 8 | GE 3T Discovery MR750 | | | | |
| 9 | GE 3T Discovery MR750 | | | | |
| 10 | GE 3T Discovery MR750 | TX 13C clamshell (GE)<br>RX 1H/13C endorectal coil (GE)<br>or a 32-channel head array coil with integrated birdcage transmit coil | 1H/13C endorectal coil | 8 M 13C urea phantom | center frequency was calibrated using 8 M 13C-urea standard |
| 11 | GE 3T Discovery MR750 | TX 13C clamshell (GE)<br>RX 16-channel bilateral phased array (RAPID)<br>or custom-built 32-channel 13C coil | | | |
| 12 | GE 3T Discovery MR750 | dual-tuned 1H/13C quadrature head coil (RAPID) | dual-tuned 1H/13C quadrature head coil (RAPID) | 8 M 13C urea attached to the ear defenders worn by the subject | 13C transmit gain (TG) and center frequency (f0) were set using a13C enriched urea phantom |
| 13 | Siemens 3T Biograph mMR | TX 13C clamshell<br>RX two (anterior and posterior) 7-channel 1H/13C receive phased array coils (RAPID) | TX 1H body coil<br>RX two (anterior and posterior) 7-channel 1 h/13C receive phased array coils (RAPID) | | |
| 14 | GE 3T Discovery MR750 | TX 13C clam-shell coil (GE)<br>RX Two 4-channel 13C surface coils arrays (GE)<br>or a 32-channel head array coil with integrated birdcage transmit coil | body coil | A head-shaped phantom containing unenriched ethylene glycol doped with 17 g/L (0.29 M) NaCl | center frequency and calibrate power (TG) with a non-slice selective 90° or 180° RF pulse, B1+ map of head phantom |
| 15 | Siemens 3T Biograph mMR | TX 13C clamshell RX two (anterior and posterior) 7-channel 1 h/13C receive phased array coils (RAPID) | TX 1H body coil<br>RX two (anterior and posterior) 7-channel 1 h/13C receive phased array coils (RAPID) | reference phantom containing 1 ml of 8M 13C-urea | The 13C receiver bandwidth was centered using a reference phantom |
| 16 | GE 3T Discovery MR750 | 8-channel 13C RX / clamshell TX<br>32-channel 13C RX / Volume TX<br>8/24-channel 1H/13C RX / Volume TX | body coil or dual-tuned 13C/1H hardware | A head-shaped phantom containing unenriched ethylene glycol doped with 17 g/L (0.29 M) NaCl | TG was calibrated using a a non-slice selective 90 pulse |
| 17 | GE 3T Discovery MR750 | custom 13C head coil | | | |
| 18 | GE 3T Discovery MR750w | TX 13C clamshell<br>RX 1H/13C endorectal coil (GE) | 1H body coil, 1H/13C endorectal coil (GE) | | |
| 19 | GE 3T Discovery MR750 | eight-channel 13C breast coil (RAPID) | 1H body coil + dedicated eight-channel phased array receive-only breast coil | 13C-labeled 8 M urea | 13C-labeled 8 M urea sample (Sigma-Aldrich), positioned adjacent to the tumor-containing breast, was used to set the 13C transmit gain and center frequency |
| 20 | GE 3T Discovery MR750 | TX volume coil, RX 8 CH or 32 CH | | | |
| 21 | GE 3T Discovery MR750 | | | | |
| 22 | Siemens 3T Tim Trio | TX 2-channel transmit, RX 8-channel surface-receive array (RAPID) | 6 channel flexible 1H receive array | 13C urea phantom | A [13C]urea fiducial marker strapped on top of the coil was used to calibrate the 13C center frequency |
| 23 | GE 3T Discovery MR750 | TX + RX custom surface coil with figure-eight configuration | 16-channel abdominal array (GE Healthcare) | | |
| 24 | GE 3T Discovery MR750 | eight-channel 13C breast coil (RAPID) | before 13C measurement 8-channel RX only breast coil | 8 M 13C urea phantom | |
| 25 | GE 3T Discovery MR750 | TX 13C clamshell (GE)<br>RX Two 4-channel 13C surface coils arrays (GE) | TX 1H body coil<br>RX 4-channel paddle receive coil | 13C urea phantom | Initial pre-scan frequency and power calibration were performed on 13C urea phantom attached outside the receive coil, which was removed before pyruvate injection. Real-time 13C frequency and power calibration and triggered upon bolus arrival. |
| 26 | GE 3T Discovery MR750 | TX 13C clamshell (GE)<br>RX 1H/13C endorectal coil (GE) | | | |
| 27 | GE 3T Discovery MR750 | | | | |
| 28 | GE 3T Discovery MR750 | TX birdcage RX 24-channel array (RAPID) or TX 13C clamshell (GE), RX Two 4-channel 13C surface coils arrays (GE) | body coil | | Immediately following imaging, a non-localized spectrum was acquired to confirm center frequency, for heart studies Integrated bolus tracking, center frequency calibration, and B1+ calibration were performed |
| 29 | GE 3T Discovery MR750w | Nested-design 1H/13C: quadrature 13C TX, and 8-channel 13C RX array (Clinical MR Solutions) | Nested-design 1H/13C: quadrature 1H TX/RX (Clinical MR Solutions) | ethylene glycol (used w/ small animal coil) | |
| 30 | GE 3T Discovery MR750w | TX Helmholtz<br>RX 8-channel receive array | body coil | | |
| 31 | GE 3T Discovery MR750w | Nested-design 1H/13C: quadrature 13C TX, and 8-channel 13C RX array (Clinical MR Solutions) | Nested-design 1H/13C: quadrature 1H TX/RX (Clinical MR Solutions) | Gd-doped 0.4-M spherical [13C]$HCO_3$– phantom (diameter = 18 cm) | single-voxel 1H point-resolved spectroscopy (PRESS) shimming up to 1st order, 13C transmit power pre-calibrated with bicarbonate phantom |
| 32 | GE 3T Discovery MR750w | Nested-design 1H/13C: quadrature 13C TX, and 8-channel 13C RX array (Clinical MR Solutions) | Nested-design 1H/13C: quadrature 1H TX/RX (Clinical MR Solutions) | | |
| 33 | GE 3T Discovery MR750 | 8-ch 13C RX / clamshell TX, 32-ch 13C RX / Volume TX | | A head-shaped phantom containing unenriched ethylene glycol doped with 17 g/L (0.29 M) NaCl, and 8M 13C-urea phantom | transmit power calibrated on a head-shaped phantom, frequency calibration with urea phantom |
| 34 | GE 3T Discovery MR750 | home-built single-tuned TX/RX 13C birdcage coil using same support base as 1H coil | 1H body coil;<br>8-channel neurovascular receive array (Invivo) | | |
| 35 | Siemens 3T Tim Trio | TX 2 channel, RX 8 channel surface-receive array (RAPID) | 1H body coil | | |
| 36 | GE 3T Discovery MR750w | 2-loop coil (20 cm diameter) TX/RX (PulseTeq)<br>flexible quadrature TX and 8 channel RX coil (Clinical MR Solutions) | 1H body coil | saline-loaded dimethyl silicone | 13C CF from 1H CF; B1+ estimated using dimethyl silicone phantoms prior to volunteer imaging |
| 37 | GE 3T Discovery MR750 | TX clamshell, RX 8-ch paddle (GE) | | | real-time CF and B1+ calibration |
| 38 | GE 3T Discovery MR750w | 2-loop coil (20 cm diameter) TX/RX (PulseTeq) | | | 13C CF from 1H CF |
| 39 | GE 3T Discovery MR750w | Nested-design 1H/13C: quadrature 13C TX, and 8-channel 13C RX array (Clinical MR Solutions) | Nested-design 1H/13C: quadrature 1H TX/RX (Clinical MR Solutions) | | |
| 40 | GE 3T Discovery MR750 | single-tuned quadrature birdcage coil as a transceiver | 32-channel head coil | urea | CF and B1+ from urea next to subject |
| 41 | GE 3T Discovery MR750 | birdcage TX + 32-channel custom-built receiver (UCSF-MGH)/ 24-channel RX (RAPID) OR custom-built volume TX/RX birdcage coil (UCSF-MGH) | | head-shaped ethylene glycol | B1+ on ethylene glycol phantom prior to subject |
| 42 | GE 3T Discovery MR750 | TX/RX custom figure-8 surface coil | commercial 32-channel torso array or a 16-channel flex array | urea integrated in 13C coil; 25-cm diameter cylindrical ethylene glycol phantom | B1+ on urea + B1 correction using B1 map from ethylene glycol phantom |
| 43 | GE 3T Discovery MR750 | 8-channel 13C breast coil (RAPID) | 1H body coil for 13C session part; 8ch RX breast array | 8-M urea | |
| 44 | GE 3T Discovery MR750 | 8-channel 1H/24-channel 13C phased array RX with an 8-rung low-pass 13C volume TX coil (RAPID) | | | 13C CF from 1H CF |
| 45 | GE 3T Discovery MR750w | 2-loop coil (20 cm diameter) TX/RX (PulseTeq) | 1H body coil | | |
| 46 | GE 3T Discovery MR750 | TX clamshell, RX 8-ch paddle (RAPID) | 32-channel cardiac array coil (GE) after respositioning | | |
| 47 | GE 3T Discovery MR750 | 1H/13C endorectal RX coil (RAPID) | 1H/13C endorectal RX coil (RAPID) | | |
| 48 | GE 3T Discovery MR750 | | | | |
| 49 | GE 3T Discovery MR750 | dual-tuned 1H/13C quadrature TX/RX head coil (RAPID) | dual-tuned 1H/13C quadrature TX/RX head coil (RAPID) | 8 M 13C urea attached to the ear defenders worn by the subject | 13C transmit gain (TG) and center frequency were set using a13C enriched urea phantom |
| 50 | GE 3T Discovery MR750w | Nested-design 1H/13C: quadrature 13C TX, and 8-channel 13C RX array (Clinical MR Solutions) | Nested-design 1H/13C: quadrature 1H TX/RX (Clinical MR Solutions) | 0.4-M bicarbonate sphere | 13C CF from 1H CF; B1+ from separate phantom scan |
| 51 | GE 3T Discovery MR750 | TX/RX volume head coil (UCSF/MIT or PulseTeq); volume TX + 24-channel RX (RAPID) | | gadolinium-doped natural abundance dimethyl silicone 16-cm sphere (for B1+ mapping) | 13C CF from 1H CF |
| 52 | GE 3T Discovery MR750 | TX/RX Helmholtz loop-pair 13C coil (PulseTeq) or TX clamshell + 16-channel RX array (RAPID) | 8-channel cardiac RX array and/or body coil | bicarbonate sphere | B1+ with Bloch-Siegert on bicarbpmate sphere close to patient; 13C CF from 1H CF |
| 53 | GE 3T Discovery MR750 | 1H/13C dual-tuned birdcage TX/RX coil (PulseTeq) | 1H/13C dual-tuned birdcage TX/RX coil (PulseTeq) | Natural abundance glycerol head phantom | B1+ with a 90◦ hard pulse on glycerol phantom (presumably prior to patient); 13C CF from 1H CF |
| 54 | GE 3T Discovery MR750 | 1H/13C TX/RX head coil (RAPID) | 12-channel head coil (GE) | | |
| 55 | GE 3T Discovery MR750 | dual-tuned 1H/13C quadrature TX/RX head coil (RAPID) | dual-tuned 1H/13C quadrature TX/RX head coil (RAPID) | | high order shimming; Bloch-Siegert for B1+ and CF |
| 56 | GE 3T Discovery MR750 | TX 13C clamshell + RX 1H/13C endorectal coil (GE) | 1H body coil for TX | urea phantom inside endorectoal coil | B1+ calibration on urea phantom; real-time CF calibration |
| 57 | GE 3T Discovery MR750 | fabricated flexible 8-ch 13C RX array + TX 13C clamshell (RAPID) | 1H body coil for TX | natural abundance ethylene glycol (for phantom experiments) | 13C CF from 1H CF; B1+ with Bloch-Siegert but not clear what signal source (23Na?) |
| 58 | GE 3T Discovery MR750 | 1H/13C endorectal RX coil (RAPID) | 1H/13C endorectal RX coil (RAPID) | | |
| 59 | Siemens 3T Biograph mMR | TX clamshell + 1H/13C endorectal RX coil (RAPID) | | 8M urea phantom inside endorectoal coil | 13C CF and B1+ calibration on urea phantom |
| 60 | GE 3T Discovery MR750 | 8 ch1H/24 ch 13C phased array receiver with an 8-rung low-pass 13C volume transmit coil (RAPID) | assuming 1H body coil TX | | |
| 61 | GE 3T Discovery MR750 | TX 13C clamshell + 1H/13C endorectal coil (GE) | 1H body coil for TX + 4ch torso array (combined with endo coil) | urea phantom inside endorectoal coil | 13C CF and B1+ calibration on urea phantom |
| 62 | GE 3T Discovery MR750 | flexible quadrature TX and 8 channel RX coil (Clinical MR Solutions) | 1H body coil | 27.8 cm-diameter spherical dimethyl silicon phantom for B1+ mapping | built-in autoshimming (in subset manual linear shim in ROI); B1+ from separate phantom scan (90deg hard pulse); 13C CF from 1H CF |
| 63 | GE 3T Discovery MR750w | volume excitation TX coil (clamshell?) + 8-channel paddle RX (GE) | | head-shaped phantom containing natural abundance ethylene glycol (for phantom experiments) | |

**Supporting Table S2**: Acquisition methods reported in HP $^{13}$C-pyruvate human study papers surveyed. Note that some papers reported multiple acquisition methods. The entries are blank when the methods were not reported in the manuscript or supplementary/supporting materials. If some values were not reported, they are denoted by a "?". In some cases, the methods were ambiguous and have been inferred from the cited paper or its references, which are *italicized* and followed by "?". "~" indicates a range of parameter values were reported. N/A = Not applicable.
For the Start Time, some studies reported a single value, but it was not clear whether this was relative to the start of injection, end of injection, or other time, and these are shown as a single value and are *italicized*.
*Abbreviations*: EPSI = Echo-planar spectroscopic imaging; SPSP = spectral-spatial; EPI = Echo-planar imaging; IDEAL = iterative decomposition with echo asymmetry and least-squares estimation; CSI = chemical shift imaging; bSSFP = balanced steady-state free-precession; RR = time interval between heart beats

| Ref | Acquisition Method (Categorical) | Acquisition Methods | Spatial resolution | Coverage | Start Time [sec] | Temporal Resolution [sec or RR intervals] | Acquired time points (start : interval : end) | Anatomical Target |
|---|---|---|---|---|---|---|---|---|
| 1 | MRS/I | 1D EPSI | 10 mm | 18 cm x 36-60 mm slice | Start of injection+0 | 3 | 3 s | Prostate |
| | MRS/I | 2D EPSI (dynamic) | 10 mm | 8 cm x 18 cm x 12-40 mm slice | Start of injection+5 | 5 | 5 s | Prostate |
| | MRS/I | 2D EPSI (single timepoint) | 7mm | 8.4 cm x 8.4 cm x 10 - 20 mm slice | Start of injection+25~33 | N/A | 12 s | Prostate |
| | MRS/I | 3D EPSI (single timepoint) | 7 mm x 7 mm x 7-15mm | 8.4 cm x 12.6 cm x 12 cm | Start of injection+25~33 | N/A | 8 - 12 s | Prostate |
| 2 | Metabolite-specific Imaging | 2D multi-slice spiral, SPSP excitation (dynamic) | 8.8 mm x 8.8 mm x 10 mm | 6 cm in slice direction | End of injection | 3 RR | Acquired 3 time points over 18 cardiac cycles | Heart |
| 3 | MRS/I | *EPSI?* | | | | | | Prostate |
| 4 | Metabolite-specific Imaging | 3D EPI | 15mm x 15mm x 15mm | 72cm x 72xm x 72cm | End of injection+10 | 6 | 10s after saline flush : 6s : 60s | Abdomen/Pelvis |
| 5 | MRS/I | 2D EPSI (dynamic) | 10-12 mm x 10-12 mm x 15-20 mm | 16-20 cm x 16-20 cm x 15-20 mm slice | End of injection+0 | 4.3 | 4.3 s | Brain |
| 6 | MRS/I | Slab dynamic 2D EPSI (dynamic) | 3 cm slice<br>15-20 mm x 15-20 mm x 20-30 mm slice | N/A<br>18-20 cm x 27-36 cm x 2-3 cm slice | End of injection+5 | 3 | 5s after end of injection : 3 s : end of injection + 5s+ 39*3s | Brain |
| 7 | MRS/I | 3D EPSI (dynamic) | 8 mm x 8 mm x 8 mm | | *End of injection+5?* | 2 | ? : 2 s : ?<br>36-s window<br>" starting 5s after injection" | Prostate |
| 8 | MRS/I | 3D EPSI (dynamic) | 8mm x 8mm x 8mm | 9.6cm x 9.6xm x 12.8cm | End of injection+5 | 2 | 5s after saline flush : 2s : 42s<br>(21 timepoints) | Prostate |
| 9 | Metabolite-specific Imaging | EPI SPSP excitation (dynamic) | | | | 2 | | Prostate |
| 9 | MRS/I | EPSI (dynamic) | | | | | | Prostate |
| 10 | Metabolite-specific Imaging | 2D EPI multi-slice (dynamic) | 8 mm x 8 mm x 8 mm<br>15 mm x 15 mm x 20 mm | 12.8 cm x 12.8 cm x 12.8 cm<br>24 cm x 24 cm x 16 cm | End of injection+5 | 2 | 5 s after end of injection: 2 s : end of injection+5s+20*2s | Methods |
| 11 | MRS/I | 2D EPSI (dynamic) | 1.8cm x 1.8cm x 2cm | ? x 28.8cm x 2cm | | 3 | ? : 3s : 60s (20 timepoints) | Abdomen/Pelvis |
| | Metabolite-specific Imaging | 2D EPI (dynamic) | 1.5 x 1.5cm x 2cm | 24cm x 24cm x 16cm | | 3 | ? : 3s : 60s (20 timepoints) | Brain |
| 12 | Chemical Shift Encoding | 2D IDEAL spiral , multi-slice (dynamic) | 12 mm x 12 mm x 30 mm | 24 cm x 24 cm x 3 cm | End of injection+10 | 4 | 10s after end of injection : 4 s : ? | Brain |
| 13 | MRS/I | 2D CSI (dynamic) | 7.5 mm x 7.5 mm x 30 mm | 12 cm x 12 cm x 3 cm slice | *25* | 20 | 25 s : 20 s : 245 s | Kidney |
| 14 | Metabolite-specific Imaging | 2D EPI multi-slice, SPSP excitation (dynamic) | 15 mm x 15 mm x 20 mm slice | ? x ? x 16 cm slice | End of injection+5 | 3 | 5s after end of injection : 3 s : ? | Brain |
| 15 | MRS/I | 2D CSI (dynamic) | 16 mm x 16 mm x 30 mm | 16 cm x 16 cm x 3 cm | *25* | 10 | 25 s : 10 s : 120 s | Breast |
| 16 | Metabolite-specific Imaging | 2D EPI multi-slice, SPSP excitation (dynamic) | 2-8 cm^3 | 24 cm x 24 cm x variable | End of injection+5 | 3 | 5s after end of flush : 3 s : 60 s<br>(20 time points) | Brain |
| 17 | Metabolite-specific Imaging | 3D EPI (dynamic, dual-echo) | 15 mm x 15 mm x 15 mm | 24 cm x 24 cm x 36 cm | End of injection+0 | 5 | End of injection: 5 s : end of injection + 60s | Brain |
| 18 | MRS/I | 2D EPSI (dynamic) | 10 mm x 10 mm x 15 mm | 16 cm x 16 cm x 15 mm slice | Start of injection+5 | 4.9 | 4.9 s | Prostate |
| 19 | Chemical Shift Encoding | 2D IDEAL spiral (2-3 slices) | Reconstructed to 2 mm x 2mm x 30 mm slice | | *12 or End of injection+12?* | 4 | *? "delay = 12 s"* | Breast |
| 20 | Metabolite-specific Imaging | 2D EPI multi-slice (dynamic) | 12-15 mm x 12-15 mm x 15-20 mm | ? x ? x 12-16 cm slice | End of injection+5 | 3 | 5s after end of injection: 3s : end of injection+5s+39*3s | Brain |
| 21 | MRS/I | Slab dynamic | | NA | End of injection+0 | 1 | 25 s : 1 s : 55 s | Pancreas |
| | MRS/I | 2D CSI (single time point) | 15 mm x 15 mm x 20 mm | 24 cm x 24 cm x 20 mm slice | *35* | N/A | 35s | Pancreas |
| 22 | MRS/I | pulse-acquire | 10 mm slice | Coil FOV x 10 mm slice | *0* | 1 RR | 0s : 1 RR interval : up to 240 s | Heart |
| 23 | MRS/I | 2D EPSI (dynamic) | 1.2-1.5cm x 1.2-1.5cm x 2-3cm | ? x ? x 2-3cm | Start of injection+5 | 3 | ? : 3s : 60s | Abdomen/Pelvis |
| 24 | Chemical Shift Encoding | IDEAL spiral | | | | | | Breast |
| 25 | Metabolite-specific Imaging | Lactate imaging: 3D bSSFP, stack of spirals (dynamic)<br>Pyruvate/alanine | 15 mm x 15 mm x 21 mm | 69 cm x 69 cm x 33.6 cm<br>45 cm x 45 cm x 33.6 cm | Detected bolus arrival+6 | 3.5 | 6s after bolus arrival : 3.5 s : ? | Kidney |
| | | imaging: 2D GRE multi-slice, spiral readout (dynamic) | | | | | | |
| 26 | Metabolite-specific Imaging | 2D EPI multi-slice, SPSP excitation (dynamic) | 8 mm x 8 mm x 8 mm slice | 12.8 cm x 12.8 cm x 11.2 cm | | 2 | ?: 2 s : ? | Prostate |
| 27 | MRS/I | 2D EPSI (dynamic) | 1.2-2cm x 1.2-2cm x 1.5-4cm slice | | | 2~5 | ? : 2-5s : (20-24 time points) | Methods |
| 28 | Metabolite-specific Imaging | 2D EPI multi-slice, SPSP excitation (dynamic) | 7.5 mm x 7.5 mm x 15 mm slice (pyruvate)<br>15 mm x 15 mm x 15 mm slice (lactate and bicarbonate) | 24 cm x 24 cm x 12 cm | End of injection+5 | 3 | 5s after end of flush : 3 s : 60 s (20 time points) | Brain |
| | Metabolite-specific Imaging | 2D spiral Gated multi-slice, SPSP excitation (dynamic) | | 2.1 cm slice x 5 slices | | 3 RR | | Heart |
| 29 | MRS/I | 2D spiral CSI single-slice (dynamic) | 15 mm x 15 mm x 2-3 cm slice | 24 cm x 24 cm x 2-3 cm | *3* | 5 | ? : 5 s : ? | Brain |
| 30 | MRS/I | Slab spectra (dynamic) | 10-cm slab | 10-cm slab | | 2.8-3.8 (*3 RR?*) | ? : 2.8-3.8 s : 3.7-5 min (80 time points) | Heart |
| 31 | MRS/I | 2D spiral CSI single-slice (dynamic) | 15 mm x 15 mm x 30 mm | 24 cm x 24 cm x 3 cm | *5* | 5 | ? : 5 s : ? | Brain |
| 32 | MRS/I | Spiral CSI (dynamic) | 15 mm x 15 mm x 25-30 mm | 24 cm x 24 cm x 2.5-3 cm | *5* | 5 | 5s after start of injection : 5 s : 90 s | Brain |
| 33 | Metabolite-specific Imaging | 2D EPI multi-slice, SPSP excitation (dynamic) | 3.38-4.5 cm^3 | 24 cm x 24 cm x 8 slices | End of injection+0~5 | 3 | 0-5 s after end of flush: 3 s : 60 s (20 time points) | Brain |
| | MRS/I | EPSI (dynamic) | 8 cm^3 | | End of injection+0~5 | 3 | 0-5 s after end of flush: 3 s : 72 s (24 time points | Brain |
| 34 | Metabolite-specific Imaging | 3D EPI with SPSP excitation (dynamic) | 15 mm x 15 mm x 15 mm | 24 cm x 24 cm x 36 cm | *0* | 5 | 0 s : 5 s : 60 s | Brain |
| 35 | Metabolite-specific Imaging | 2D hybrid-shot spiral multi-slice, SPSP excitation | 10 mm x 10 mm x 20 mm<br><br>6 mm x 6 mm x 10 mm | ? x ? x 60 mm<br><br>38.4 cm x 38.4 cm x 12 cm | End of injection+22 | | Data acquired over 9 heartbeats, starting 22 s after injection<br>Data acquired over 36 heartbeats, starting 22 s after injection | Heart |
| 36 | Metabolite-specific Imaging | 2D spiral, SPSP excitation (dynamic) | 6.1 mm x 6.1 mm x ? | 40 cm x 40 cm x ? | End of injection+0 | 6 RR | end of injection: 6 RR : 50 s | Heart |
| 37 | Metabolite-specific Imaging | 2D EPI multi-slice, SPSP excitation (dynamic) | 15 mm x 15 mm x 21 mm | | Detected bolus arrival+6 | 4 | 6s after bolus arrival : 4s : ? | Kidney |
| | Metabolite-specific Imaging | 3D bSSFP (Lactate only) | 15 mm x 15 mm x 21 mm | | Detected bolus arrival+6 | 4 | 6s after bolus arrival : 4s : ? | Kidney |
| 38 | Metabolite-specific Imaging | 2D spiral multi-slice, SPSP excitation (dynamic) | 16 mm x 16 mm x 30 mm | 40 cm x 40 cm x 30 mm | *25* | 2 RR | 25s after start of injection : 2 RR : 25s + 16 timepoints. | Heart |
| 39 | MRS/I | Slab spectra (dynamic) | 100 mm slab | 100 mm slab | | 3 | ? : 3 s : 240 s | Muscle |
| | Metabolite-specific Imaging | Multi-echo spiral with metabolite-specific SPSP excitation | ? x ? x 80 mm slice thickness | | *33* | N/A | | Muscle |
| 40 | Metabolite-specific Imaging | 2D EPI multi-slice, SPSP excitation (dynamic) | 15 mm x 15 mm x 15 mm | ? x ? x 12 cm | End of injection | 3 | 0s : 3s : 57s | Brain |
| 41 | Metabolite-specific Imaging | 2D EPI multi-slice, SPSP excitation (dynamic) | 15 mm x 15 mm x 15 mm | 24 cm x 24 cm x 12 cm | End of injection+5 | 3 | 0 : 3 s : 60 s | Brain |
| 42 | MRS/I | 2D EPSI with multi-band SPSP excitation (dynamic) | 12 mm x 12 mm x 12-30 mm | ? x ? x 12-30 mm | End of injection+5 | 3 | 0 : 3 s : 60 s | Liver |
| 43 | Metabolite-specific Imaging | Metabolite-specific SPSP excitation and spiral readout. One patient-IDEAL spiral | 5 mm x 5 mm x ? | 20 cm x 20 cm x ? | | 4 | ? : 4 s : ? | Breast |
| | Chemical Shift Encoding | Metabolite-specific SPSP excitation and spiral readout. One patient-IDEAL spiral | 5 mm x 5 mm x ? | 20 cm x 20 cm x ? | | 4 | ? : 4 s : ? | Breast |
| 44 | Metabolite-specific Imaging | Metabolite specific EPI with SPSP excitation (dynamic) | 15 mm x 15 mm x 15 mm | 24 cm x 24 cm x 12 cm | Start of injection+5 | 3 | 0 s (5s after end of flush): 3 s : 60 s | Brain |
| 45 | Metabolite-specific Imaging | Metabolite-specific SPSP excitation and spiral readout (dynamic) | 10 mm x 10 mm x 30 mm | 40 cm x 40 cm x 3 cm | *25* | 1 RR | Started at 25s after start of injection. One acquisition per RR interval. | Heart |
| 46 | Chemical Shift Encoding | IDEAL spiral multi-slice (dynamic) | 17 mm x 17 mm x 30 mm | 34 cm x 34 cm x 17 cm (5 slices with 5 mm slice gap) | *12* | 4 | ? : 0.5 s (?) : ? "starting 12 s after the injection of HP-pyruvate" | Kidney |
| 47 | Chemical Shift Encoding | IDEAL spiral (dynamic) | 12.5 mm x 12.5 mm x ? | 20 cm x 20 cm x ? | | 4 | 0 s : 4 s : 80 s | Prostate |
| 48 | | | | | | | | Prostate |
| 49 | Chemical Shift Encoding | IDEAL spiral | 6 mm x 6 mm x 30 mm | 24 cm x 24 cm x 3 cm | *10* | 4 | 10 s after injection : 4 s : 70 s | Brain |
| 50 | MRS/I | Slab MRS (dynamic) | 15 mm slab | 15 mm slab | | 3 | ? : 3 s : 240 s | Brain |
| 51 | Metabolite-specific Imaging | EPI with SPSP excitation (dynamic) | 7.5-15 mm x 7.5-15 mm x 15 mm | | End of injection+2 | 3 | 2 s after end of injection: 3 s : ? | Brain |
| 52 | Metabolite-specific Imaging | Metabolite-specific SPSP with spiral readout (dynamic) | 13.3 mm x 13.3 mm x 30 mm | 40 cm x 40 cm x 3 cm | | 3 RR | Acquired 40 images of each metabolite with a time resolution of 3 heart beats (images in diastole) | Heart |
| 53 | Metabolite-specific Imaging | Metabolite-specific SPSP with spiral readout (dynamic) | 0.875 mm x 0.875 mm x 20 mm (pyruvate). 1.75 mm x 1.75 mm x 20 mm (lactate/bicarbonate) | ? x ? x 12 cm | End of injection+0 | 2 | : 2s : | Brain |
| 54 | Chemical Shift Encoding | IDEAL spiral | 6 mm x 6 mm x 30 mm | 24 cm x 24 cm x 3 cm | | 4 | 0 s : 4 s : 60 s | Brain |
| 55 | MRS/I | 2D CSI multi-slice (single time point) | 20 mm x 20 mm x 20 mm | 20 cm x 20 cm x 10 cm | *22~27* | N/A | 22-27 s "after injection" | Brain |
| 56 | Metabolite-specific Imaging | Metabolite-specific 2D GRE (pyruvate/lactate) Metabolite-specific 3D-bSSFP (urea) (dynamic) | 7 mm x 7 mm x 11.6 mm | Pyruvate/lactate: 22.4 cm x 22.4 cm x 11.6 cm<br>Urea: 21.7 cm x 21.7 cm x 18.56 cm | End of injection+8 | 2.6 | 8s after end of flush : 2.6 s : 52 s | Prostate |
| 57 | Metabolite-specific Imaging | Metabolite-specific stack of spirals (dynamic) | 7 x 7 x 15 mm (pyruvate) 14x14x15 mm (lactate/alanine/bicarbonate) | 28 x 28 x 12 cm | Start of injection+0 | 1.232 | 0 s after end of pyruvate injection, before flush : 1.232 : ? | Brain |
| 58 | Chemical Shift Encoding | | | | | 4 | | Prostate |
| 59 | Chemical Shift Encoding | multi-echo bSSFP (dynamic) | 11.3 mm x 11.3 mm x 10 mm | 90 mm x 90 mm x 80 mm | Start of injection+0 | 6 | 0 s after end of injection: 6 s : 204 s | Prostate |
| 60 | Metabolite-specific Imaging | 2D EPI with SPSP excitation multi-slice (Dynamic) | 7.5 mm x 7.5 mm x 15-20 mm (pyruvate). 15mm x 15 mm x 15-20 mm (lactate/bicarbonate) | 24 cm x 24 cm x 1.5-2 cm (pyruvate). 48 cm x 48 cm x 1.5-2cm (lactate/bicarbonate) | End of injection+0~5 | 3 | 0 s (5s after end of flush OR immediately after end of flush) : 3s : 60s | Brain |
| 61 | Metabolite-specific Imaging | Metabolite-specific EPI with SPSP excitation (dynamic) | 6.5-8mm x 6.5-8mm x 8 mm | 10.4-12.8 cm x 10.4-12.8 cm x 11.2 cm | Start of injection+10 | | 0 s (10s after end of flush): | Prostate |
| 62 | MRS/I | 2D EPSI (dynamic) | 20-22 mm x 20-22 mm 20-30 mm | 32-35.2cm x 36-39.6cm x 2-3cm | End of injection+5 | 3 | 5s after end of flush: 3 s : 60 s | Liver |
| | Metabolite-specific Imaging | EPI (dynamic) | 20 mm x 20 mm x 20 mm | 32 cm x 32 cm x 2 cm | End of injection+5 | 3 | 5s after end of flush: 3 s : 60 s | Liver |
| 63 | MRS/I | 2D EPSI (dynamic) | 12.5 mm x 12.5 mm x 15 mm | 20 cm x 20 cm x 1.5 cm | Start of injection+0 | 4.3 | 0s: 4.3 s: 58 timepoints | Methods |

**Supporting Table S3**: Metrics reported in HP $^{13}$C-pyruvate human study papers surveyed.

*Abbreviations*: pyr = pyruvate; lac = lactate; ala = alanine; bic = bicarbonate; AUC = area-under-curve; tC = total Carbon-13 signal; PK = pharmacokinetic; kPL, kPA, kPB, kLP are kinetic rates; ve and vb are voxel extracellular and vascular fractions, respectively; T1eff = T1 effective; TTP = time to peak; NAWM = normal appearing white matter; NAB = normal appearing brain

*Khegai PK model reference*: Khegai O, Schulte RF, Janich MA, et al. Apparent rate constant mapping using hyperpolarized [1-(13)C]pyruvate. NMR Biomed. 2014;27(10):1256-1265. doi:10.1002/nbm.3174

*Inputless PK model reference*: Larson PEZ, Chen HY, Gordon JW, et al. Investigation of analysis methods for hyperpolarized 13C-pyruvate metabolic MRI in prostate cancer patients. NMR Biomed. 2018;31(11):e3997. doi:10.1002/nbm.3997; Hyperpolarized-MRI-Toolbox. https://github.com/LarsonLab/hyperpolarized-mri-toolbox, doi:10.5281/zenodo.1198915

| Reference | Metrics | Values Reported | Model Parameters |
|---|---|---|---|
| 1 | kPL mean<br>lac/pyr | 0.009 – 0.013 sec-1 | Fit T1pyr, T1lac<br>No reverse conv |
| 2 | AUC Pyr<br>AUC Lac<br>AUC Bic | ~95-140 (plot)<br>~35-65<br>~30-80 | |
| 3 | kPL max (Prostate tumor) | 0.007 - .025 sec-1 | model not specified |
| 4 | AUC pyr<br>AUC lac<br>AUC ala<br>AUC Bic | ~50-13500 (colorbar)<br>~50-920<br>~50-670<br>~50-550 | |
| 5 | AUC pyr<br>AUC lac<br>TTP pyr<br>TTP lac<br>kPL, full 2-comp model | ~0.0-0.7<br>~0.0-0.1<br>11.7 +/- 1.9<br>23.0 +/- 1.3<br>0.12 (0.08-0.16) s-1 | Fixed T1pyr;<br>Constrained T1lac, ve |
| 6 | SNR lac max (normal)<br>SNR bic max (normal)<br>SNR lac max (tumor)<br>SNR bic max (tumor)<br>Lac/pyr (normal)<br>Bic/pyr (normal)<br>Lac/pyr (tumor)<br>Bic/pyr (tumor) | 28.0-79.1<br>9.2-34.1<br>4.0-51.6<br>1.5-5.8<br>0.18-0.98<br>0.07-0.37<br>0.30-0.58<br>0.02-0.08 | |
| 7 | kPL<br>SNR pyr (last, mean)<br>SNR lac (last, mean)<br>SNR total 13C<br>SNR total 13C (normal tissue) | ~.005-.0.020 sec-1 (colorbar)<br>104, 45.2<br>10.7, 6.1<br>51.3<br>48.2 | Fit kLP=0; Fit t1, T1s equal |
| 8 | Inputless kPL max<br>Inputless kPL mean<br>AUC lac/AUC pyr<br>kPL with input<br>TTP Lac<br>Mean lac time<br>Mean pyr time<br>SNR total Pyr, max<br>SNR total Lac, max | 0.023 (0.009 - 0.049)<br>0.009 (0.003-0.018)<br><br><br><br><br>29.0 (24.8-32.3) sec<br>628.8 (189.2-1415.0)<br>99.4 (49.5-195.5) | Fit T1pyr, T1LAC; kLP=0<br>Fixed T1lac; kLP=0<br>Fixed T1lac; kLP=0<br>From PK model w/input<br>Fixed T1LAC, fit input |
| 9 | kPL<br>Unconstrained<br>Constrained | <br>~0.00-0.02 s-1<br>~0.001-0.0075 s-1 | Fitting with vascular input function(defined)<br>kLP=vb=0; |
| 10 | Dynamic images<br>AUC pyr, lac, bic (brain)<br>Peak SNR pyr, lac, bic (brain)<br>TTP pyr, lac, bic (brain) | <br>0-1, 0-0.28, 0-0.07<br>396-447, 16-30, 6-8<br>3-9, 6-12, 9-15 | |
| 11 | Pyr AUC (brain)<br>Lac AUC (brain)<br>Bic AUC (brain)<br>Lac/Pyr AUC ratio (brain)<br>Bic/Pyr AUC ratio (brain)<br>Lac/Pyr (liver tumor)<br>Lac/Pyr (normal liver) | ~0-0.75<br>~0-0.22<br>~0-0.12<br>~0-1<br>~0-0.5<br>0.200-0.306<br>0.070-0.108 | |
| 12 | AUC: pyr (norm to max pyr)<br>Lac (norm to max pyr)<br>Bic (norm to max pyr)<br>AUC ratios: lac/pyr<br>Bic/pyr<br>Bic/lac<br>kPL<br>kPB | <br><br><br>0.18±0.04 - 0.25±0.08<br>0.04±0.02 - 0.08±0.05<br>0.20±0.02 – 0.33±0.1<br>0.008 - 0.024 s-1<br>0.002 – 0.003 s-1 | |
| 13 | Pyr<br>lac<br>Lac/pyr | ~5-35<br>~1-5.5<br>~0.1-0.55 | <br>First timepoint (20s) |
| 14 | Pyr SNR<br>Lac SNR<br>Bic SNR | | |
| 15 | Pyr (t=25s)<br>Lac (t=25s) | ~3-16<br>~0.5-2.5 | |
| 16 | kPL (NAWM)<br>kPB (NAWM)<br>kPL (lesion)<br>kPL-lesion/kPL-NAWM | 0.014-0.047 s-1<br>0.0033-0.011 s-1<br>0.012-0.51<br>0.76-1.30 | “inputless” PK model |
| 17 | Lac AUC<br>Bic AUC<br>Lac Z-Score<br>Bic Z-Score | Image<br>Image<br>~-3-2.6<br>~-2.3-3 | Z-score as the standard deviation from average |
| 18 | Pyr TTP<br>Lac TTP<br>Max Lac/tC<br>Cumulative normalized Lac | 11.2-33.5 s<br>~20-61s<br>0.1485-0.6383<br>0.0885-0.4200 | |
| 19 | Lac/Pyr AUC ratio<br>Total Pyr SNR<br>Total Lac SNR<br>kPL | ~0-0.6<br>~10-160<br>~0-50<br>~0-0.45 s-1 | Two-way PK; equal T1s; vb=0 |
| 20 | kPL,<br>kLP,<br>kPB | 0.009-0.05<br><br>~0-0.009 | (models defined)<br>Fit T1 assuming equal for metabolites; vb=0 |
| | Mean arrival time<br>Max SNR pyr, lac, bic | 6-11s<br>101-437,13-79,11-27 | |
| 21 | Lac/Pyr<br>Ala/Pyr<br>Ala/Lac | 0.36<br>0.12<br>0.33 | |
| 22 | Bic/Pyr AUC ratio<br>Lac/Pyr AUC ratio<br>Bic/Lac AUC ratio<br>Ala/Pyr AUC ratio | 0.0016-0.0184<br>0.0516-0.1053<br>0.02-0.30<br>0.0317-0.0474 | |
| 23 | kPL<br>SNR tC | 0.013-0.026 s-1<br>19.5-290.3 | “inputless” PK model |
| 24 | kPL<br>Lac/Pyr AUC ratio | ~0.15-0.25<br>~0.30-0.50 | Khegai model |
| 25 | Lac/Pyr AUC ratio<br>AUC Images | ~0-2.4 | |
| 26 | kPL’ | ~0.005-0.035 | 2 physical compartment model with arterial input function<br>T1Pyr, T1Lac assumed |
| 27 | Metabolite images<br>SNR<br>kPL | <br><br>0.0261-0.0369 s-1 | “inputless” PK model |
| 28 | Pyr SNR<br>Lac SNR<br>Bic SNR<br>Peak Pyr SNR<br>Peak Lac SNR<br>Peak Bic SNR<br>Pyr SNR (left ventricle)<br>Lac SNR<br>Bic SNR<br>Pyr AUC<br>Lac AUC<br>Bic AUC | ~0-87<br>~0-15<br>~0-6<br>288.8<br>14.8<br>6.3<br>~0-906<br>~0-50<br>~0-14<br>~1.6-162<br>~0.3-30<br>~0.1-4.0 | |
| 29 | Pyr Signal Intensity<br>Lac SI | Images,<br>Dynamic Curves | |
| 30 | Bic/tC<br>Lac/tC<br>Ala/tC | 0.032-0.038<br>0.061-0.448<br>0.045-0.051 | |
| 31 | Bic/tC<br>Lac/tC | 0.025-0.059<br>0.162-0.236 | |
| 32 | Lac/tC AUC (Tumor)<br>Lac/tC AUC (NAB)<br>Bic/tC AUC (Tumor)<br>Bic/tC AUC (NAB) | 0.160-0.327<br>0.148-0.268<br>0.062-0.111<br>0.077-0.124 | |
| 33 | Pyr AUC<br>Lac AUC<br>Bic AUC<br>Lac/Pyr AUC Ratio<br>Pyr AUC SNR (EPI)<br>Lac AUC SNR (EPI)<br>Bic AUC SNR (EPI) | ~0-88<br>~0-76<br>~0-26<br>~0-1.8<br>785-1070<br>175-243<br>64-83 | |
| | Pyr AUC SNR (EPSI)<br>Lac AUC SNR (EPSI)<br>Bic AUC SNR (EPSI) | 4750-352000<br>522-72000<br>304-15200 | |
| 34 | Lac Z-score | -7 | |
| 35 | Pyr AUC<br>Lac AUC<br>Bic AUC | ~10-80<br>~2-12<br>~1-6 | |
| 36 | Bic SNR<br>Pyr AUC<br>Lac AUC<br>Bic AUC | 6.8 – 14.9<br>Figure<br>Figure<br>Figure | |
| 37 | Pyr SNR<br>Lac SNR<br>Ala SNR<br>Pyr AUC<br>Lac AUC<br>Ala AUC<br>Lac/pyr AUC ratio<br>Ala/pyr AUC ratio | 75 ± 49<br>13 ± 5.8<br>4.2 ± 2.0<br>Figure<br>Figure<br>Figure<br>~0.5 – 1.6<br>0.48 – 1.72 | |
| 38 | Pyr T2*<br>Lac T2*<br>Bic T2*<br>Pyr SNR<br>Lac SNR<br>Bic SNR | 109 – 129 ms<br>41 – 44 ms<br>64 ms<br>~100 – 250<br>~10 – 50<br>~10 – 100 | |
| 39 | Total carbon AUC<br>Pyr/tC AUC ratio<br>Lac/tC AUC ratio<br>Ala/tC AUC ratio<br>Bic/tC AUC ratio<br>Ala/lac AUC ratio<br>Bic/lac AUC ratio | <br>0.41 – 0.66<br>0.18 – 0.42<br>0.11 – 0.22<br>0.002 – 0.004<br>0.28 – 0.72<br>0.004 – 0.036 | tC adjusted for polarization |
| 40 | kPL<br>Lac/Pyr AUC ratio<br>Pyr relative cerebral blood flow (rCBF)<br>Pyr mean transit time (MTT) | ~0.10-0.30<br>~0.2-0.75<br>~0.5-2.5<br>~12.5-25 s | ‘Inputless’ PK model |
| 41 | kPL<br>kPB<br>Peak Pyr AUC SNR<br>Peak Lac AUC SNR<br>Peak Bic AUC SNR | Mean:0.0133-0.0221 s-1<br>Mean:0.0022-0.0093 s-1<br>160-393<br>17-56<br>7-13 | Model not specified; T1s fit |
| 42 | Pyr SNR fold-increase<br>Lac SNR fold-increase<br>Ala SNR fold-increase<br>kPL | <br><br><br>0.027 – 0.073 s-1 | ‘Inputless’ PK model |
| 43 | Pyr SNR<br>Lac SNR<br>Lac/pyr AUC ratio<br>kPL | 16.5 – 19.7<br>5.6 – 7<br>0.28 – 0.34<br>0.0064 – 0.0079 s-1 | PK model not specified (SVD recon) |
| 43 | Pyr SNR<br>Lac SNR | 16.5 – 19.7<br>5.6 – 7 | PK model not specified (SVD recon) |
| | Lac/pyr AUC ratio<br>kPL | 0.28 – 0.34<br>0.0064 – 0.0079 s-1 | |
| 44 | *Global*<br>Pyr SNR fold-increase with denoising<br>Lac SNR fold-increase with denoising<br>Bic SNR fold-increase with denoising<br><br>*NAWM*<br>kPL<br>kPB<br>Bic/lac AUC ratio<br><br>*Non-enhancing lesion:*<br>kPL<br>kPB<br>Bic/lac AUC ratio<br><br>*Contrast enhancing lesion*<br>kPL<br>kPB<br>Bic/lac AUC ratio | <br>3.9 +/- 0.8 (NOT SNR..)<br>3.1 +/- 1.1<br>5.1 +/- 2.2<br><br><br>~0.015 – ~0.033 s-1<br>~0.0035 – ~0.005 s-1<br>~0.15 – ~0.275<br><br><br>~0.01 – ~0.04 s-1<br>~0.0025 s-1<br>~0.1 – ~0.15<br><br><br>~0.0175 – ~0.02 s-1<br>~0.002 s-1<br>~0.075 – ~0.1 | <br>'inputless' PK model |
| 45 | Bic/tC AUC ratio<br>Lac/tC AUC ratio<br>Ala/tC AUC ratio<br>Bic/lac AUC ratio<br>Bic /Ala AUC ratio | 0.04 – 0.22<br>0.143 – 0.17<br>0.07 – 0.13<br>0.3 – 1.4<br>0.63 – 2.6 | |
| 46 | *Tumor*<br>Pyr SNR<br>Lac SNR<br>Lac/pyr AUC ratio<br>kPL<br><br>*Contralateral*<br>Pyr SNR<br>Lac SNR<br>Lac/pyr AUC ratio<br>kPL | <br>26.7 (16.8-61.3)<br>5.7 (1.9-9.6)<br>0.13 – 0.35<br>0.0022 – 0.0152 s-1<br><br><br>30.1 (14.8-64-7)<br>3.5 (2-9.3)<br>0.14<br>0.0043 s-1 | PK with Khegai model. Two-site exchange with T 1eff |
| 47 | *Tumor*<br>Total carbon SNR<br>Pyr SNR<br>Lac SNR<br>kPL | <br>27 – 112<br>15 – 70<br>6 – 40<br>0.003 – 0.018 s-1 | PK with Khegai model. Two-site exchange with T 1eff |
| 48 | kPL | 0 – 0.027 s-1 | Not reported |
| 49 | Lac AUC<br>Pyr AUC<br>Lac/(lac+pyr) AUC ratio | <br><br>~0.21 | |
| 50 | Lac/tC<br>Bic/tC<br>Ala/tC<br>Bic/lac<br>Pyr-hydrate/pyr | 0.21 – 0.24<br>0.065 – 0-091<br>0.002 – 0.026<br>0.3 – 0.38<br>0.048 – 0-067 | Ratios from time-averaged spectra<br><br>Three-site exchange PK model, fitting of T1's and k's. |
| | Lac/pyr<br>Bic/pyr<br>Pyr-hydrate/tC<br><br>Upslope lac / upslope pyr<br>Upslope bic / upslope pyr<br>TTP pyr<br>TTP lac<br>TTP bic<br><br>kPL<br>kPB | ~0.38<br>~0.12<br>~0.04<br><br>~0.06<br>~0.015<br>~17 s<br>~27 s<br>~34 s<br><br>~0.015<br>~0.005 | |
| 51 | Pyr SNR<br>Lac SNR<br><br>kPL<br>Lac/pyr AUC ratio<br>Lac Z-score | 10.9<br>13.8<br><br>~0.015 – 0.025 s-1<br>~0.2 – 0.67 | 'Inputless' PK model |
| 52 | Pyr SNR<br>Lac SNR<br>Ala SNR<br>Bic SNR<br>Lac TTP<br>Ala TTP<br>Bic TTP<br>Pyr TTP<br>Pyr first order moment<br><br>kPL<br>kPB<br>kPA | 150 – 290<br>7 – 19<br>9 – 17<br>7 – 12<br>13 – 16 s<br>6 – 11 s<br>13 – 17 s<br>6 – 3 s<br>7 - 17.8 s<br><br>0.011 – 0.02 s-1<br>0.004 – 0.012 s-1<br>0.005 – 0.016 s-1 | 'Inputless' PK model |
| 53 | kPL<br>kPB<br>Lac/bic AUC ratio<br>Pyr mean transit time | 0.014 – 0.023 s-1<br>~0.0065 s-1<br>0.4 – 0.045<br>~15 s | 'Inputless' PK model |
| 54 | *Normal brain*<br>Pyr AUC / peak pyr<br>Lac AUC / peak pyr<br>Bic AUC / peak pyr<br>Lac/pyr AUC ratio<br>Bic/pyr AUC ratio<br>kPL<br>kPB<br><br>*Glioblastoma multiforme*<br>Pyr AUC / peak pyr<br>Lac AUC / peak pyr<br>Bic AUC / peak pyr<br>Lac/pyr AUC ratio<br>Bic/pyr AUC ratio | <br>0.14 – 0.24<br>0.04 – 0.09<br>0.009 – 0.027<br>0.36 +/-0.06<br>0.1 +/- 0.03<br>0.0165 +/- 0.0073 s-1<br>0.0024 +/- 0.001 s-1<br><br><br>0.19 – 0.41<br>0.05 – 0.13<br>0.006 – 0.022<br>0.34 +/- 0.06<br>0.06 +/- 0.03 | AUC ratios noise-corrected<br>PK with Khegai model. Two-site exchange with T 1eff |
| | kPL<br>kPB | 0.0161 +/- 0.0057 s-1<br>0.0017 +/- 0.0013 s-1 | |
| 55 | Pyr/max tC<br>Lac/max tC<br>Bic/max tC<br>Lac/pyr<br>Bic/pyr<br>Lac/bic | 0.63 ± 0.09<br>0.28 ± 0.09<br>0.08 ± 0.02<br>0.48 ± 0.23<br>0.13 ± 0.03<br>3.68 ± 1.21 | |
| 56 | Pyr AUC<br>Lac AUC<br>Lac/pyr AUC ratio | ~30-150<br>~10-25<br>~0.1-0.5 | |
| 57 | | | |
| 58 | *Tumor*<br>Total carbon SNR<br>Pyr SNR<br>Lac SNR<br>kPL | <br>42.7 – 94.1<br>22.6 – 61.2<br>9.2 – 23.1<br>0.005 – 0.018 s-1 | PK with Khegai model. Two-site exchange with T 1eff |
| 59 | lac/pyr AUC ratio<br><br>kPL | 0.33 +/- 0/12 (tumor), 0.15+/- 0.10 (healthy)<br>0.038 /- 0.014 s-1 (tumor), 0.011 +/- 0.007 s-1 (healthy) | unidirectional PK model, T1pyr and kPL estimated |
| 60 | Lac/pyr AUC ratio<br>kPL | 0.24 – 0.42<br>0.01 – 0.015 s-1 | ‘Inputless’ PK model |
| 61 | kPL | 0.0198 – 0.041 s-1 | ‘Inputless’ PK model |
| 62 | Pyr SNR<br>Lac SNR<br>Ala SNR<br><br>Relative pyr TTP<br><br>Pyr AUC<br>Lac AUC<br>Ala AUC<br>kPL<br>kPA | <br><br><br><br><br><br><br><br><br>0.0033 – 0.019 s-1<br>0.00073 – 0.012 s-1 | ’Inputless’ PK model |
| 63 | SNR (spatial uniformity) | Images | NA |